\documentclass[11pt]{iopjournalfixed}

\makeatletter

\renewcommand{\title}[1]{%
  {\exhyphenpenalty=10000
   \hyphenpenalty=10000
   \fontsize{18}{21}\selectfont
   \centering
   \textsf{#1}\par}%
  \suppressfloats[t]%
}

\renewcommand{\author}[1]{%
  \vspace{5mm}%
  {\fontsize{10}{12}\selectfont
   \centering
   \if@anonymous
     Author list removed for anonymity%
   \else
     #1%
   \fi
   \par}%
  \vspace{3mm}%
}

\renewcommand{\affil}[1]{%
  {\fontsize{8}{10}\selectfont
   \centering
   \if@anonymous\else #1\fi
   \par}%
}

\renewcommand{\email}[1]{%
  \vspace*{12pt}%
  {\fontsize{8}{10}\selectfont
   \centering
   \if@anonymous
   \else
     \textbf{E-mail addresses:} #1%
   \fi
   \par}%
  \vspace{3mm}%
}

\makeatother

\usepackage[english]{babel}
\usepackage[utf8]{inputenc}
\usepackage[T1]{fontenc}
\usepackage{amsmath}
\usepackage{amssymb}
\usepackage{indentfirst}
\usepackage{amsthm}

\usepackage[numbers,sort&compress]{natbib}

\numberwithin{equation}{section}
\newtheorem{theorem}{Theorem}[section]
\newtheorem{corollary}[theorem]{Corollary}

\runningtitle{Higher-curvature gravity}
\runningauthors{Melgarejo, Molano, Ram{\'i}rez}
\hypersetup{
  pdftitle={Perturbative vacuum constraints in higher-curvature gravity:
    Schwarzschild deformations and strong-field observables},
  pdfauthor={Jonathan Ram{\'i}rez, Gustavo Melgarejo, and Daniel Molano},
  pdfsubject={Four-dimensional higher-curvature gravity and black-hole perturbations},
  pdfkeywords={modified gravity, higher-curvature gravity, black holes,
    perturbation theory, Gauss--Bonnet gravity, black-hole shadows, Wald entropy}
}

\begin{document}
\pagestyle{plain}

\title{Perturbative vacuum constraints in higher-curvature gravity:
Schwarzschild deformations and strong-field observables}

\author{Gustavo Melgarejo$^1$\orcid{0000-0002-7811-5970}, Daniel Molano$^{2,3,4}$\orcid{0009-0007-0509-6565}, and Jonathan Ramírez$^1$\orcid{0009-0005-6018-0124}}

\affil{$^1$Universidade do Estado do Rio de Janeiro (UERJ), CEP 20550-013, Rio de Janeiro, RJ, Brazil}

\affil{$^2$Escuela Colombiana de Ingeniería Julio Garavito, Departamento de Ciencias Básicas, Bogotá, Colombia}
\affil{$^3$Pontificia Universidad Javeriana, Departamento de Economía, Bogotá, Colombia}
\affil{$^4$Universidad Sergio Arboleda, Departamento de Matemáticas, Bogotá, Colombia}
\affil{Corresponding author: Daniel Molano.}

\email{ga.melgarejoc@gmail.com; da.molanom@uniandes.edu.co; ramirez.jsg@gmail.com}
\begin{abstract}
Higher-curvature terms modify the gravitational action, but they do notnecessarily generate new vacuum geometries on the branch perturbatively connected to general relativity. We develop a first-order framework for static and spherically symmetric vacuum black holes in four-dimensional metric theories with gravitational Lagrangian
\(\mathcal L_{\mathrm{grav}}=R/2+\lambda\Psi(R,X,Y)\), where
\(X=R_{\mu\nu}R^{\mu\nu}\) and
\(Y=R_{\mu\nu\rho\sigma}R^{\mu\nu\rho\sigma}\). On the branch analytic
in the coupling, purely Ricci-based terms \(\Psi(R,X)\) that are analytic
in their curvature arguments and satisfy \(\Psi(0,0)=0\) do not source
nontrivial vacuum deformations of a Ricci-flat background under the
regularity and boundary assumptions adopted here. Riemann-dependent
terms, by contrast, can generate such deformations because the
background Kretschmann scalar remains nonzero. In areal-radius gauge, we
derive model-independent first-order expressions for the horizon shift,
ISCO, epicyclic frequencies, periapsis advance, photon sphere, critical
shadow impact parameter, Wald entropy, and Hawking temperature.

We apply the framework to the power-law Gauss--Bonnet model
\(\Psi_\eta=\ell_\lambda^{-2+4\eta}\mathcal G^\eta\), treating
noninteger powers as phenomenological parametrizations of nonlinear
curvature effects, and obtain a closed first-order solution on the
Schwarzschild-connected branch. The perturbations decay at infinity for
\(\eta>1/2\), but a standard fixed-ADM interpretation requires
\(\eta>2/3\); at \(\eta=2/3\), the sourced \(1/r\) term mixes with the
asymptotic mass mode. For positive effective coupling, the horizon,
ISCO, photon sphere, and critical shadow impact parameter increase for
\(2/3<\eta<1\) and decrease for \(\eta>1\). On the fixed-ADM branch, the
mass, Wald entropy, and Hawking temperature satisfy
\(dM=T_H\,dS_{\rm W}\) through first order under variations at fixed
fundamental couplings. At \(\eta=1\), the linear four-dimensional
Gauss--Bonnet term leaves the local geometry and geodesic observables
unchanged while producing a constant topological entropy shift. Finally,
an Event Horizon Telescope-inspired shadow-size criterion provides a
conservative estimate of the first-order sensitivity to the effective
dimensionless coupling. This is a geometric consistency test rather than
a complete observational constraint.

\end{abstract}

\section{Introduction}

General relativity (GR) remains the most successful classical theory of
gravity. It has passed tests ranging from Solar System experiments and binary-pulsar timing to gravitational-wave observations and horizon-scale observations of
black holes
\cite{Will2014,BertottiIessTortora2003,KramerEtAl2021,Abbott2016,EHT2019M87,EHT2022SgrAMetric}. Nevertheless, spacetime
singularities, the absence of a perturbatively renormalizable quantum
theory of GR, the cosmological constant problem, and the possible
gravitational origin of the dark sector continue to motivate
gravitational theories beyond Einstein's theory
\cite{Penrose1965,HawkingEllis1973,tHooftVeltman1974,Weinberg1989,
CLIFTON20121,CapozzielloDeLaurentis2011,SotiriouFaraoni2010,
DeFeliceTsujikawa2010}. From an effective-field-theory viewpoint,
higher-curvature corrections to the Einstein--Hilbert action are expected,
particularly in high-curvature regimes where ultraviolet or semiclassical
effects may become relevant
\cite{Stelle1977,Buchbinder1992,Donoghue1994,
NojiriOdintsovOikonomou2017}.

Multimessenger astronomy provides a natural arena in which to test such
corrections. Compact-binary gravitational waves probe a dynamical strong-field
regime, while very-long-baseline interferometry by the Event Horizon Telescope
(EHT) accesses ring and shadow-scale observables associated with the photon
region of supermassive compact objects
\cite{Abbott2016,EHT2019M87,EHT2022SgrA,EHT2022SgrAMetric}. Small departures
from the Schwarzschild or Kerr geometries can therefore affect the innermost
stable circular orbit (ISCO), epicyclic frequencies, relativistic precession,
photon spheres, critical shadow scales, and horizon thermodynamics. Controlled perturbative frameworks are essential for distinguishing genuine predictions of a modified theory from artifacts caused by extrapolating a finite-order expansion beyond its regime of validity, as well as from solutions lying on disconnected branches.

\begin{figure}[h]
    \centering
    \includegraphics[width=0.7\linewidth]{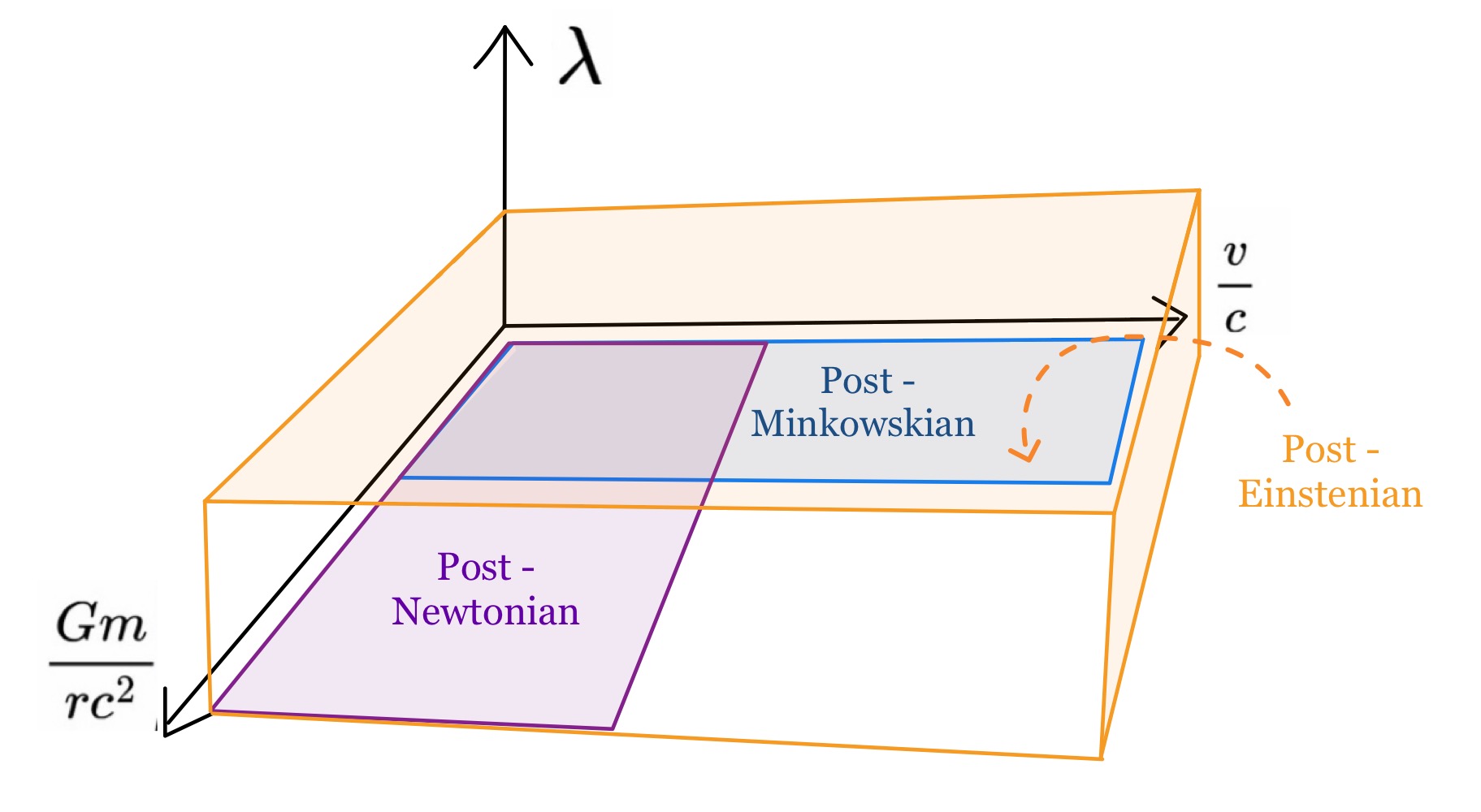}
    \caption{
   Schematic representation of different perturbative regimes in gravity. The plane $\lambda=0$ corresponds to GR, where the post-Newtonian and post-Minkowskian expansions describe weak-field limits. For small but nonzero $\lambda$, weak-field modified-gravity tests probe deviations from GR near the Newtonian or Minkowskian sectors. The orange region represents the post-Einsteinian perturbative regime considered in this work: small deviations from a strong-field GR background, such as a Schwarzschild black hole, induced by higher-curvature corrections.}
    \label{fig:post-einstein}
\end{figure}

The philosophy adopted here is analogous to that of post-Newtonian and
post-Minkowskian expansions. The former compares relativistic gravity with
Newtonian gravity in the weak-field, slow-motion regime, whereas the latter
expands about Minkowski spacetime without assuming slow motion
\cite{Will2014,Blanchet2024}. Here GR itself is the background theory: we ask
how a modified metric theory departs from a fully relativistic GR solution as
its coupling is switched on. We call this a \emph{post-Einsteinian}
expansion in theory space.
Figure~\ref{fig:post-einstein} illustrates the hierarchy of limits and locates
the present strong-field expansion within the space of gravitational
approximations.

A central difficulty in modified gravity is that its field equations are
generically nonlinear, of higher differential order, and difficult to solve
exactly. It can therefore be unclear whether a solution is continuously
connected to a GR spacetime or belongs to a distinct or nonanalytic branch;
quadratic gravity provides explicit examples of additional black-hole solutions
\cite{LuPerkinsPopeStelle2015}. Previous work established a perturbative
criterion for comparing GR vacuum solutions with those of Ricci-based modified
metric theories
\cite{MolanoBargueno2025,MolanoVillalbaCastanedaBargueno2020}. In particular,
for the one-parameter family
\begin{equation}
f_\lambda(R,X)=R+2\lambda\Psi(R,X),
\qquad X=R_{\mu\nu}R^{\mu\nu},
\end{equation}
provided that $\Psi(R,X)$ is analytic in a neighborhood of
$(R,X)=(0,0)$ and that the solution admits an analytic expansion in
$\lambda$ about a Ricci-flat GR background, such as the Schwarzschild
solution, the perturbative vacuum equations reduce order by order to
their GR counterparts. Any different vacuum solution must then be
disconnected from the GR branch or nonanalytic at \(\lambda=0\). Thus, not
every formal modification of the Lagrangian produces a genuine perturbative
deformation of a GR vacuum solution.

The present work extends that program to theories depending on the three
curvature scalars \(R\), \(X=R_{\mu\nu}R^{\mu\nu}\), and
\(Y=R_{\mu\nu\rho\sigma}R^{\mu\nu\rho\sigma}\). This extension is essential:
although \(R\) and \(R_{\mu\nu}\) vanish on a Ricci-flat background such as
Schwarzschild, the Riemann tensor and the Kretschmann scalar do not. Nonlinear
dependence on Riemann-sector invariants can consequently source first-order
metric perturbations, producing coupling-analytic Schwarzschild-like
deformations absent from purely Ricci-based models.

This paper is organized into three main parts. The first part (Section \ref{sec:fourth-order-gravity}) develops the general theoretical framework. We introduce fourth-order higher-curvature gravity, define the perturbative construction on the extended manifold $\mathcal{N}=\mathcal{M}\times\mathbb{R}$, and explain how tensors defined on a modified-gravity spacetime can be pulled back to the GR background. This allows the GR solution and the modified solution to be compared in a fixed perturbative gauge. Within this setting, we revisit the Ricci-based constraints and show why invariants that vanish on a Ricci-flat background do not source analytic vacuum corrections, whereas Riemann-dependent invariants may do so.

The second part (Section \ref{sec:phenomenology}) derives a model-independent dictionary between the first-order metric perturbations and physically relevant black-hole quantities. Working in the static and spherically symmetric case, we obtain general expressions for the shift of the Killing horizon, the innermost stable circular orbit, the radial epicyclic frequency, the periapsis-precession correction, the photon-sphere radius, and the critical shadow impact parameter. These expressions provide a direct way to translate a perturbative modified-gravity solution into potentially observable strong-field diagnostics. We also compute the first-order Wald entropy correction, separating the contribution due to the perturbed horizon area from the explicit higher-curvature contribution to the entropy functional \cite{Wald1993,IyerWald1994}.

The third part (Section \ref{sec:power-law-gb}) applies the general formalism to a
power-law Gauss--Bonnet model,
\begin{equation}
    \Psi_{\eta}(\mathcal{G})\propto \mathcal{G}^{\eta}.
\end{equation}
The Gauss--Bonnet invariant is theoretically distinguished for several reasons.
It is the quadratic Lovelock density and therefore represents a geometrically
preferred combination of curvature-squared terms
\cite{Lovelock,Fernandes_2022}. In four dimensions, its integral is proportional, up to boundary terms, to the Euler characteristic of the manifold and does not modify the local metric field equations \cite{Lanczos1938}. In dimensions larger than four, however, the same linear combination contributes dynamically while preserving second-order field equations. Moreover,
the Gauss--Bonnet density arises naturally among the leading curvature-squared corrections to low-energy string effective actions
\cite{Zwiebach1985,Green_Schwarz_Witten_2012,Ortin_2004}, while nonlinear
functions of $\mathcal{G}$ have been widely considered as effective
parametrizations of gravitational dynamics beyond GR
\cite{NojiriOdintsov2005,DeFeliceTsujikawaCosmology2009,
NojiriOdintsov2011,NojiriOdintsovOikonomou2017}. Perturbative corrections to
Schwarzschild observables in \(f(\mathcal G)\) gravity were studied in
Ref.~\cite{DeFeliceTsujikawaSolarSystem2009}, and exact static solutions of
models including \(R+\sqrt{\mathcal G}\) were constructed in
Ref.~\cite{MyrzakulovSebastianiZerbini2013}. A recent analysis
derived asymptotic black-hole corrections and shadow observables in
\(f(R,\mathcal G)\) gravity \cite{Nashed2026}. Here we instead obtain a closed
first-order radial solution on the Schwarzschild-connected branch and a common
dictionary for orbital, optical, and thermodynamic quantities.

The Gauss--Bonnet sector is also particularly well suited to the perturbative
problem considered here. On a Ricci-flat Schwarzschild background,
\begin{equation}
    R^{(0)}=0,
    \qquad
    R_{\mu\nu}^{(0)}=0,
    \qquad
    \mathcal{G}^{(0)}
    =
    R_{\mu\nu\rho\sigma}^{(0)}
    R^{(0)\mu\nu\rho\sigma}
    \neq 0,
\end{equation}
so that $\mathcal{G}$ provides a natural mechanism for evading the perturbative
rigidity of purely Ricci-based vacuum theories. The power-law family offers an
analytically tractable laboratory in which $\eta=1$
provides a nontrivial consistency check, whereas nonlinear values of $\eta$
generate genuine local sources for Schwarzschild-like first-order deformations.
For noninteger $\eta$, the model should be understood as a phenomenological
effective parametrization of nonlinear curvature corrections rather than as a
unique ultraviolet completion. We solve the first-order field equations, determine the admissible
parameter regimes, and compute the corresponding corrections to the
horizon, ISCO, epicyclic frequencies, periapsis precession, photon
sphere, and critical shadow impact parameter. We then use the
shadow-scale correction to obtain conservative EHT-inspired consistency
estimates and present a synthetic geometric visualization comparing the
Schwarzschild shadow scale with its power-law Gauss--Bonnet deformation.
Finally, we compute the Wald entropy and Hawking temperature and verify
first-law consistency.
The goal of this work is therefore twofold. First, we provide a systematic post-Einsteinian perturbative framework for comparing extended gravitational theories with GR in the strong-field regime. Second, we show explicitly how this framework can be used to extract observable and thermodynamic information from a concrete higher-curvature model. In doing so, we aim to clarify which modified-gravity corrections can be analytically connected to GR and how such corrections may be constrained by current and future multimessenger observations.


\section{Fourth-order higher-curvature gravity and perturbation theory}
\label{sec:fourth-order-gravity}

In this section, we introduce the higher-curvature framework and the
perturbative construction used throughout the paper. We first consider metric
theories whose gravitational Lagrangian depends on the curvature invariants
\(R\), \(R_{\mu\nu}R^{\mu\nu}\), and
\(R_{\mu\nu\rho\sigma}R^{\mu\nu\rho\sigma}\). We then formulate the comparison
between a general-relativistic background and a one-parameter family of
modified-gravity solutions on an extended manifold. This construction allows
us to distinguish between Ricci-based contributions, which do not generate new
analytic vacuum deformations around Ricci-flat backgrounds, and genuinely
Riemann-dependent contributions, which can provide nonvanishing first-order
sources.

\subsection{Fourth-order higher-curvature gravity}
\label{sec:higher-curvature-framework}

One route toward constructing gravitational theories beyond general
relativity involves relaxing the assumptions underlying Lovelock's theorem.
In four spacetime dimensions, Lovelock's theorem establishes that general
relativity, possibly supplemented by a cosmological constant, is the unique
local metric theory constructed from the metric and its derivatives whose
field equations are of second differential order
\cite{Lovelock}. Once this requirement is relaxed, the Einstein--Hilbert
action can be supplemented by higher-curvature invariants. Such terms arise
naturally as counterterms in quantum field theory in curved spacetime, in
effective-field-theory expansions of gravity, and in ultraviolet-motivated or
string-inspired descriptions
\cite{Stelle1977,BirrellDavies1982,Buchbinder1992,Donoghue1994,
Zwiebach1985}. They have also been widely studied in connection with
inflation, late-time cosmology, and possible strong-field deviations from the
Schwarzschild and Kerr geometries
\cite{Starobinsky1980,SotiriouFaraoni2010,DeFeliceTsujikawa2010,
NojiriOdintsov2011,NojiriOdintsovOikonomou2017,Fernandes_2022}.

We adopt the metric signature $(-,+,+,+)$, and the curvature convention
\begin{equation}
    R^{\rho}{}_{\sigma\mu\nu}
    =
    \partial_{\mu}\Gamma^{\rho}_{\nu\sigma}
    -
    \partial_{\nu}\Gamma^{\rho}_{\mu\sigma}
    +
    \Gamma^{\rho}_{\mu\alpha}\Gamma^{\alpha}_{\nu\sigma}
    -
    \Gamma^{\rho}_{\nu\alpha}\Gamma^{\alpha}_{\mu\sigma},
    \label{eq:riemann-convention}
\end{equation}
with $ R_{\mu\nu} = R^{\rho}{}_{\mu\rho\nu}$.

We consider metric theories constructed from the three curvature invariants
\begin{equation}
    R,
    \qquad
    X\equiv R_{\mu\nu}R^{\mu\nu},
    \qquad
    Y\equiv
    R_{\mu\nu\rho\sigma}R^{\mu\nu\rho\sigma}.
    \label{eq:curvature-invariants}
\end{equation}
This choice keeps the Ricci-scalar, Ricci-tensor, and Riemann-tensor contributions
explicit and provides a convenient framework for identifying which
higher-curvature terms remain nonzero on a Ricci-flat background.

The gravitational action is written as\footnote{We use $\kappa^{2}=8\pi G$ and $c=1$.}
\begin{equation}
    S[g,\psi_m]
    =
    \frac{1}{2\kappa^{2}}
    \int d^{4}x\,\sqrt{-g}\,
    f_{\lambda}(R,X,Y)
    +
    S_m[g,\psi_m],
    \label{eq:actiongen}
\end{equation}
where \(S_m\) denotes the matter action. The energy--momentum tensor is defined by
\begin{equation}
    T_{\mu\nu}
    =
    -\frac{2}{\sqrt{-g}}
    \frac{\delta S_m}{\delta g^{\mu\nu}}.
    \label{eq:stress-energy-definition}
\end{equation}

The one-parameter family of theories considered in the perturbative analysis
is normalized as
\begin{equation}
    f_{\lambda}(R,X,Y)
    =
    R
    +
    2\lambda\Psi(R,X,Y).
    \label{eq:f-lambda-definition}
\end{equation}
 The parameter \(\lambda\) is taken to be
dimensionless, while \(\Psi\) has dimensions of inverse length squared.

For a general function \(f(R,X,Y)\), we use the notation
\begin{equation}
    f_A
    \equiv
    \frac{\partial f}{\partial A},
    \qquad
    A=R,X,Y.
\end{equation}
Variation of Eq.~\eqref{eq:actiongen} with respect to the inverse metric gives
the field equations
\begin{equation}
    P_{\mu\nu}
    =
    \kappa^{2}T_{\mu\nu},
    \label{eq:field-equations-general}
\end{equation}
where
\begin{align}
P_{\mu\nu}
={}&
f_R R_{\mu\nu}
-\frac{1}{2}g_{\mu\nu}f
+2f_XR_{\mu\alpha}R^{\alpha}{}_{\nu}
+2f_YR_{\mu\alpha\beta\gamma}
       R_{\nu}{}^{\alpha\beta\gamma}
\nonumber\\
&+
g_{\mu\nu}\Box f_R
-\nabla_{\mu}\nabla_{\nu}f_R
+\Box\left(f_XR_{\mu\nu}\right)
+g_{\mu\nu}\nabla_{\alpha}\nabla_{\beta}
        \left(f_XR^{\alpha\beta}\right)
\nonumber\\
&-
2\nabla_{\alpha}\nabla_{(\mu}
        \left(f_XR_{\nu)}{}^{\alpha}\right)
+4\nabla_{\alpha}\nabla_{\beta}
        \left(f_YR_{\mu}{}^{\alpha}{}_{\nu}{}^{\beta}\right).
\label{eq:Pmunu-general}
\end{align}
This expression follows the curvature convention in
Eq.~\eqref{eq:riemann-convention}; changing the index ordering of the Riemann
tensor changes the displayed sign of its double-divergence term
\cite{Carroll:2004de}.
Here $\Box\equiv g^{\mu\nu}\nabla_{\mu}\nabla_{\nu},$
and parentheses around tensor indices denote symmetrization, $B_{(\mu\nu)}=\frac{1}{2}(B_{\mu\nu}+B_{\nu\mu})$. Since \(f_R\), \(f_X\), and \(f_Y\)
are themselves functions of curvature invariants, the metric field equations
are generically of fourth differential order.

A geometrically distinguished combination within this class is the
Gauss--Bonnet invariant,
\begin{equation}
    \mathcal G
    =
    R^{2}
    -
    4R_{\mu\nu}R^{\mu\nu}
    +
    R_{\mu\nu\rho\sigma}R^{\mu\nu\rho\sigma}
    =
    R^{2}-4X+Y.
    \label{eq:gauss-bonnet-invariant}
\end{equation}
It is the quadratic Lovelock density and has a special status among
curvature-squared combinations. In four dimensions, the integral of
\(\mathcal G\) is topological, up to boundary contributions. Consequently, a linear Gauss--Bonnet correction does not
modify the local metric field equations.

Nonlinear functions of \(\mathcal G\), by contrast, are not topological. A
family of theories of particular interest is obtained by choosing
\begin{equation}
    \Psi(R,X,Y)
    =
    \Phi(\mathcal G),
\end{equation}
so that
\begin{equation}
    f_{\lambda}(R,X,Y)
    =
    R
    +
    2\lambda\Phi(\mathcal G).
    \label{eq:f-gauss-bonnet-general}
\end{equation}
When \(\Phi(\mathcal G)\) is linear,
\(\Phi_{\mathcal G}\) is constant and the four-dimensional local contribution
vanishes. For a nonlinear function, however,
\(\Phi_{\mathcal G}\) is generally spacetime dependent, and its covariant
derivatives generate genuine local terms in the metric field equations
\cite{Lanczos1938,Zwiebach1985,Fernandes_2022}. This makes nonlinear
Gauss--Bonnet gravity a natural setting in which to study analytic
higher-curvature deformations of Ricci-flat black-hole geometries.

\subsection{Perturbative construction on the extended manifold}
\label{sec:perturbation-ricci-constraints}

We now introduce the perturbative framework used to compare a solution of GR
with a nearby solution of a modified gravitational theory. The construction is
formulated on the extended manifold
\begin{equation}
    \mathcal N
    =
    \mathcal M\times\mathbb R,
    \label{eq:extended-manifold}
\end{equation}
which represents a one-parameter family of spacetime slices
\begin{equation}
    \mathcal M_{\lambda}
    \equiv
    \mathcal M\times\{\lambda\}.
\end{equation}
The slice \(\mathcal M_0\) is identified with the GR background spacetime,
whereas each slice \(\mathcal M_{\lambda}\), with \(\lambda\neq0\), represents
a spacetime belonging to the modified theory. The family of theories is
assumed to reduce to GR when \(\lambda\rightarrow0\).\footnote{
A detailed discussion of perturbation theory on the extended manifold
\(\mathcal N=\mathcal M\times\mathbb R\), together with its different
implementations in gravitational theories, can be found in
Ref.~\cite{Molano2025PRD}. Three distinct scenarios are identified there:
perturbations within GR, perturbations within a fixed extended theory of
gravity, and the comparison between a GR background on \(\mathcal M_0\) and a
modified-gravity solution on \(\mathcal M_{\lambda}\). The present work follows
the third scenario, in which the parameter \(\lambda\) labels both the
departure from GR and the family of spacetimes being compared.
}

Let \(X^{a}\) be a smooth vector field on \(\mathcal N\), transverse to the
slices \(\mathcal M_{\lambda}\). Its flow defines a family of maps
\begin{equation}
    \mathcal X_{\lambda}
    :
    \mathcal M_0
    \longrightarrow
    \mathcal M_{\lambda}.
    \label{eq:gauge-map}
\end{equation}
The map \(\mathcal X_{\lambda}\) identifies points on the background spacetime
with points on the modified spacetime. In relativistic perturbation theory,
this identification is known as a gauge choice of the second kind.

On the background slice, we assume that the GR field equations are satisfied,
\begin{equation}
    \mathcal E_0(g_0,\tau_0)
    =
    0,
    \label{eq:background-field-equation}
\end{equation}
where \(g_0\) and \(\tau_0\) denote the background metric and
energy--momentum tensor. On each modified slice, we assume
\begin{equation}
    \mathcal E_{\lambda}
    (g_{\lambda},\phi_{\lambda},\tau_{\lambda})
    =
    0,
    \label{eq:modified-slice-equation}
\end{equation}
where \(g_{\lambda}\) is the metric on \(\mathcal M_{\lambda}\),
\(\tau_{\lambda}\) is the corresponding energy--momentum tensor, and
\(\phi_{\lambda}\) collectively denotes any additional fields that may be
present in the modified theory. The configurations studied in the remainder
of this work are vacuum configurations, for which
\begin{equation}
    \tau_{\lambda}=0.
\end{equation}

To compare tensors defined on different slices, they are pulled back to the
background manifold. For a smooth tensor field \(T\) on \(\mathcal N\), the
pullback admits the expansion
\begin{equation}
    \mathcal X_{\lambda}^{*}T
    =
    \sum_{n=0}^{\infty}
    \frac{\lambda^{n}}{n!}
    T^{(n)},
    \qquad
    T^{(n)}
    \equiv
    \left.
    \mathcal L_X^{n}T
    \right|_{\lambda=0},
    \label{eq:taylor-tensor}
\end{equation}
where \(\mathcal L_X\) is the Lie derivative along \(X^{a}\). In particular,
for the metric we define
\begin{equation}
    \bar g_{ab}
    \equiv
    \mathcal X_{\lambda}^{*}g_{\lambda ab}.
\end{equation}
Its perturbative expansion is
\begin{equation}
    \bar g_{ab}
    =
    g^{(0)}_{ab}
    +
    \lambda h_{ab}
    +
    \frac{\lambda^{2}}{2}g^{(2)}_{ab}
    +
    \mathcal O(\lambda^{3}),
    \label{eq:metric-expansion}
\end{equation}
where
\begin{equation}
    h_{ab}
    \equiv
    g^{(1)}_{ab}
\end{equation}
is the first-order metric perturbation.

The expansion in Eq.~\eqref{eq:taylor-tensor} assumes that the family of
pulled-back fields is sufficiently smooth and admits a Taylor expansion in
\(\lambda\) around \(\lambda=0\). Solutions that are disconnected from the GR
branch, or whose dependence on \(\lambda\) is nonanalytic, are not described
by this perturbative construction.

The coefficients \(T^{(n)}\) generally depend on the gauge choice
\(\mathcal X_{\lambda}\). Nevertheless, the vacuum field equations vanish on
every slice. Consequently, their perturbations are gauge invariant order by
order under the generalized Stewart--Walker result. This property is
particularly useful in the comparison between vacuum GR and vacuum
modified-gravity solutions \cite{StewartWalker1974,Bruni1997}.

\begin{figure}[tbp]
    \centering
    \includegraphics[width=0.7\linewidth]{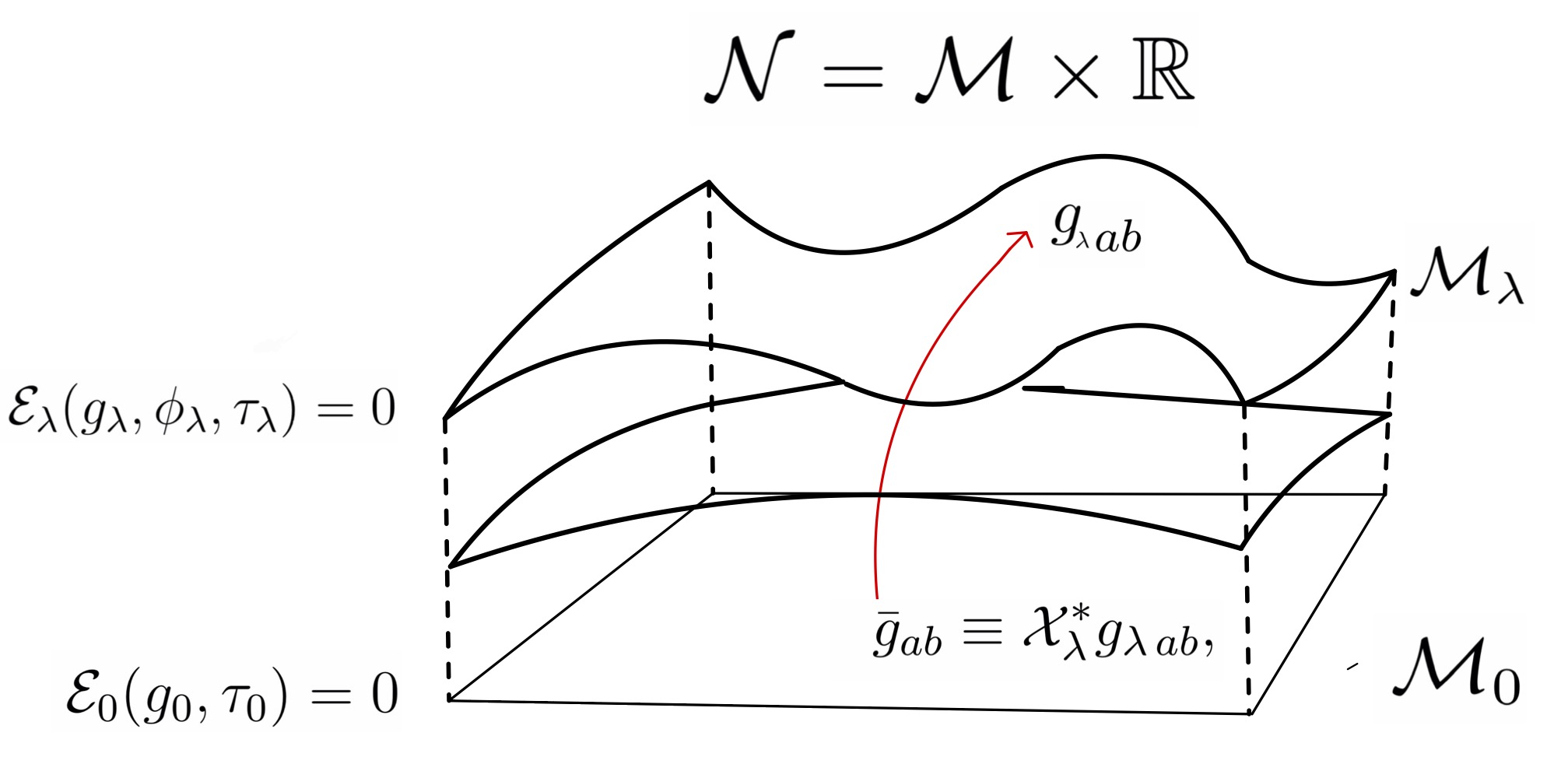}
    \caption{
    Perturbative construction on the extended manifold
    \(\mathcal N=\mathcal M\times\mathbb R\). The slice
    \(\mathcal M_0\) represents the GR background spacetime, whereas
    \(\mathcal M_{\lambda}\) represents a spacetime belonging to the modified
    theory. The pullback \(\mathcal X_{\lambda}^{*}\) maps tensors defined on
    \(\mathcal M_{\lambda}\) to \(\mathcal M_0\), allowing both geometries to
    be compared in a fixed perturbative gauge.
    }
    \label{fig:gauge-pullback}
\end{figure}

\subsection{Ricci-based rigidity and Riemann-dependent sources}
\label{sec:ricci-riemann-selection}

A central result of Ref.~\cite{MolanoBargueno2025} is that the perturbative
vacuum behavior of a modified gravitational theory depends crucially on which
curvature invariants remain nonzero on the GR background. For Ricci-based
Lagrangians, the analytic vacuum branch around a Ricci-flat background does
not generate sourced perturbative deviations from GR. We recall this result
without reproducing its inductive proof.

\begin{theorem}[Ricci-based perturbative rigidity]
\label{thm:ricci-rigidity}
Let \(g_{\lambda}\) be a one-parameter family of vacuum metrics admitting a
Taylor expansion in \(\lambda\) around \(\lambda=0\), with
\(g_0\) a Ricci-flat solution of GR. Consider a Ricci-based theory of the form
\begin{equation}
    f_{\lambda}(R,X)
    =
    R
    +
    2\lambda\Psi(R,X),
    \qquad
    X=R_{\mu\nu}R^{\mu\nu},
    \label{eq:theorem-fRX}
\end{equation}
where \(\Psi(R,X)\) is analytic in a neighbourhood of
\((R,X)=(0,0)\) and satisfies
\begin{equation}
    \Psi(0,0)=0.
\end{equation}
Then, on the analytic vacuum branch connected to \(g_0\), the perturbative
field equations coincide order by order with the perturbative Einstein
equations:
\begin{equation}
    P_{\mu\nu}^{(n)}
    =
    G_{\mu\nu}^{(n)},
    \qquad
    n\geq0.
    \label{eq:order-by-order-ricci-rigidity}
\end{equation}
Equivalently,
\begin{equation}
    R_{\mu\nu}^{(n)}=0,
    \qquad
    R^{(n)}=0,
    \qquad
    n\geq0.
    \label{eq:ricci-orders-vanish}
\end{equation}
\end{theorem}

The theorem does not require spherical symmetry or any other spacetime
symmetry. It applies to arbitrary Ricci-flat vacuum backgrounds, including
black holes and gravitational-wave spacetimes. It implies that a
Ricci-based correction cannot act as a nonzero perturbative source on the
analytic branch connected to GR. Once the asymptotic mass, time
normalization, and other boundary data have been fixed, no genuinely new
vacuum deformation remains.

Solutions of the Ricci-based theory that differ from the corresponding GR
solution may nevertheless exist. Such solutions must fall outside at least
one of the hypotheses of Theorem~\ref{thm:ricci-rigidity}. In particular, they
may belong to a branch disconnected from GR, fail to be analytic in
\(\lambda\), or arise through a singular perturbation problem in which the
higher-order equations require boundary conditions incompatible with the
naive GR limit. The latter possibility is associated with boundary-layer
behavior.

It is important to distinguish two different analyticity assumptions. The
first concerns the function \(\Psi(R,X)\) as a function of its curvature
arguments near the Ricci-flat point \((0,0)\). The second concerns the
dependence of the family of solutions \(g_{\lambda}\) on the coupling
parameter \(\lambda\). Theorem~\ref{thm:ricci-rigidity} requires both
properties. In the power-law Gauss--Bonnet model considered later,
noninteger powers are instead defined on the branch
\(\mathcal G>0\) associated with the Schwarzschild background; their
asymptotic behavior will be discussed separately in
Sec.~\ref{sec:power-law-gb}.

Two direct consequences of Theorem~\ref{thm:ricci-rigidity} will be useful.

\begin{corollary}[\(f(R)\) vacuum rigidity]
\label{cor:fR-rigidity}
Consider
\begin{equation}
    f_{\lambda}(R)
    =
    R
    +
    2\lambda\Psi(R),
\end{equation}
where \(\Psi\) is analytic near \(R=0\) and
\(\Psi(0)=0\). On the analytic vacuum branch connected to a Ricci-flat GR
background,
\begin{equation}
    P_{\mu\nu}^{(n)}
    =
    G_{\mu\nu}^{(n)},
    \qquad
    n\geq0.
\end{equation}
\end{corollary}

\begin{corollary}[Linear quadratic-curvature sector]
\label{cor:linear-quadratic-sector}
Consider the one-parameter family of four-dimensional quadratic Lagrangians
\begin{equation}
    f_{\lambda}
    =
    R
    +2\lambda\left(
    \alpha R^{2}
    +\beta R_{\mu\nu}R^{\mu\nu}
    +\gamma R_{\mu\nu\rho\sigma}R^{\mu\nu\rho\sigma}
    \right),
    \label{eq:linear-quadratic-lagrangian}
\end{equation}
with constant coefficients. Using
\begin{equation}
    Y
    =
    \mathcal G-R^{2}+4X,
\end{equation}
the Riemann-squared term can be rewritten as a combination of Ricci-based
quadratic terms plus the Gauss--Bonnet topological density. Therefore, on the
analytic vacuum branch around a Ricci-flat GR background, the perturbative
field equations reduce to those of GR.
\end{corollary}

The previous results motivate the inclusion of curvature invariants that
remain nonzero on a Ricci-flat background. For the Schwarzschild geometry,
\begin{equation}
    R^{(0)}=0,
    \qquad
    R_{\mu\nu}^{(0)}=0,
    \label{eq:ricci-flat-background}
\end{equation}
but
\begin{equation}
    R_{\mu\nu\rho\sigma}^{(0)}
    \neq0.
\end{equation}
In particular, the background Kretschmann scalar,
\begin{equation}
    Y^{(0)}
    \equiv
    R_{\mu\nu\rho\sigma}^{(0)}
    R^{(0)\mu\nu\rho\sigma},
    \label{eq:background-kretschmann}
\end{equation}
is nonvanishing. Nonlinear Riemann-dependent contributions can therefore act as effective sources for a first-order metric deformation while remaining
connected to GR through the parameter \(\lambda\).

Using the normalization introduced in
Eq.~\eqref{eq:f-lambda-definition}, the vacuum field equations can be written
as
\begin{equation}
    \mathcal E_{\mu\nu}
    \equiv
    G_{\mu\nu}
    +
    2\lambda\mathcal H_{\mu\nu}
    =
    0,
    \label{eq:modified-field-equations}
\end{equation}
where \(G_{\mu\nu}\) is the Einstein tensor and
\(\mathcal H_{\mu\nu}\) contains the contribution generated by
\(\Psi(R,X,Y)\):
\begin{align}
\mathcal H_{\mu\nu}
={}&
-\frac{1}{2}g_{\mu\nu}\Psi
+\Psi_RR_{\mu\nu}
+2\Psi_XR_{\mu\alpha}R^{\alpha}{}_{\nu}
+2\Psi_YR_{\mu\alpha\beta\gamma}
R_{\nu}{}^{\alpha\beta\gamma}
\nonumber\\
&+
g_{\mu\nu}\Box\Psi_R
-\nabla_{\mu}\nabla_{\nu}\Psi_R
+\Box\left(\Psi_XR_{\mu\nu}\right)
+g_{\mu\nu}\nabla_{\alpha}\nabla_{\beta}
   \left(\Psi_XR^{\alpha\beta}\right)
\nonumber\\
&-
2\nabla_{\alpha}\nabla_{(\mu}
   \left(\Psi_XR_{\nu)}{}^{\alpha}\right)
+4\nabla_{\alpha}\nabla_{\beta}
   \left(
   \Psi_YR_{\mu}{}^{\alpha}{}_{\nu}{}^{\beta}
   \right).
\label{eq:Hmunu-general}
\end{align}

After pulling the equations back to \(\mathcal M_0\), we expand
\begin{equation}
    \bar{\mathcal E}_{\mu\nu}
    =
    \mathcal E_{\mu\nu}^{(0)}
    +
    \lambda\mathcal E_{\mu\nu}^{(1)}
    +
    \frac{\lambda^{2}}{2}
    \mathcal E_{\mu\nu}^{(2)}
    +
    \mathcal O(\lambda^{3}).
\end{equation}
The perturbative coefficients satisfy
\begin{align}
    \mathcal E_{\mu\nu}^{(0)}
    &=
    G_{\mu\nu}^{(0)},
    \label{eq:E-zero-order}
    \\
    \mathcal E_{\mu\nu}^{(n)}
    &=
    G_{\mu\nu}^{(n)}
    +
    2n\mathcal H_{\mu\nu}^{(n-1)},
    \qquad
    n\geq1.
    \label{eq:E-orders}
\end{align}

At zeroth order, the vacuum equation is
\begin{equation}
    G_{\mu\nu}^{(0)}
    =
    R_{\mu\nu}^{(0)}
    -
    \frac{1}{2}
    R^{(0)}g_{\mu\nu}^{(0)}
    =
    0.
    \label{eq:zeroth-vacuum}
\end{equation}
The static and spherically symmetric background used below is Schwarzschild
and satisfies Eq.~\eqref{eq:ricci-flat-background}.

At first order,
\begin{equation}
    G_{\mu\nu}^{(1)}
    =
    -2\mathcal H_{\mu\nu}^{(0)},
    \label{eq:first-order-structure}
\end{equation}
so that the higher-curvature tensor evaluated on the GR background acts as an
effective source for \(h_{\mu\nu}\).

On a Ricci-flat background, every term explicitly proportional to
\(R^{(0)}\) or \(R_{\mu\nu}^{(0)}\) vanishes. In particular, the complete
\(\Psi_X\) sector vanishes because it contains the background Ricci tensor.
The remaining contribution is
\begin{align}\nonumber
\mathcal H_{\mu\nu}^{(0)}
={}&
-\frac{1}{2}
g_{\mu\nu}^{(0)}\Psi^{(0)}
+
2\Psi_Y^{(0)}
R_{\mu\alpha\beta\gamma}^{(0)}
R_{\nu}^{(0)\,\alpha\beta\gamma}+
g_{\mu\nu}^{(0)}
\Box\Psi_R^{(0)}\\
&
-
\nabla_{\mu}\nabla_{\nu}\Psi_R^{(0)}
+
4\nabla_{\alpha}\nabla_{\beta}
\left(
\Psi_Y^{(0)}
R_{\mu}^{(0)\,\alpha}{}_{\nu}{}^{\beta}
\right).
\label{eq:H-background-unsimplified}
\end{align}
Here all contractions and covariant derivatives are evaluated with the
background metric, and
\begin{equation}
    \Psi^{(0)}
    \equiv
    \Psi\left(
    R^{(0)},X^{(0)},Y^{(0)}
    \right),
\end{equation}
while
\begin{equation}
    \Psi_A^{(0)}
    \equiv
    \left.
    \frac{\partial\Psi}{\partial A}
    \right|_{
    \left(
    R^{(0)},X^{(0)},Y^{(0)}
    \right)},
    \qquad
    A=R,X,Y.
\end{equation}

The contracted Bianchi identity implies
\begin{equation}
    \nabla_{\alpha}
    R_{\mu}{}^{\alpha}{}_{\nu}{}^{\beta}
    =
    -\nabla_{\nu}R_{\mu}{}^{\beta}
    +
    \nabla^{\beta}R_{\mu\nu}.
    \label{eq:contracted-riemann-bianchi}
\end{equation}
Therefore, on a Ricci-flat background,
\begin{equation}
    \nabla_{\alpha}
    R_{\mu}^{(0)\,\alpha}{}_{\nu}{}^{\beta}
    =
    0.
\end{equation}
Expanding the last term in
Eq.~\eqref{eq:H-background-unsimplified}, the derivatives acting on the
background Riemann tensor consequently vanish, and one obtains
\begin{align}
\mathcal H_{\mu\nu}^{(0)}
={}&
-\frac{1}{2}
g_{\mu\nu}^{(0)}\Psi^{(0)}
+
2\Psi_Y^{(0)}
R_{\mu\alpha\beta\gamma}^{(0)}
R_{\nu}^{(0)\,\alpha\beta\gamma}+
g_{\mu\nu}^{(0)}
\Box\Psi_R^{(0)}
\nonumber\\
&-
\nabla_{\mu}\nabla_{\nu}\Psi_R^{(0)}
+
4R_{\mu}^{(0)\,\alpha}{}_{\nu}{}^{\beta}
\nabla_{\alpha}\nabla_{\beta}\Psi_Y^{(0)}.
\label{eq:H-background}
\end{align}
The general first-order equation is therefore
\begin{equation}
    G_{\mu\nu}^{(1)}=
g_{\mu\nu}^{(0)}\Psi^{(0)}
-4\Psi_Y^{(0)}
R_{\mu\alpha\beta\gamma}^{(0)}
R_{\nu}^{(0)\,\alpha\beta\gamma}-
2g_{\mu\nu}^{(0)}
\Box\Psi_R^{(0)}
+
2\nabla_{\mu}\nabla_{\nu}\Psi_R^{(0)}
-
8R_{\mu}^{(0)\,\alpha}{}_{\nu}{}^{\beta}
\nabla_{\alpha}\nabla_{\beta}\Psi_Y^{(0)}.
\label{eq:first-order-main}
\end{equation}

Since the Einstein tensor is
identically conserved, the higher-curvature contribution satisfies
\(\nabla^{\mu}\mathcal H_{\mu\nu}=0\) identically. At the background level,
\begin{equation}
    \nabla^{\mu}
    \mathcal H_{\mu\nu}^{(0)}
    =
    0,
    \label{eq:source-conservation}
\end{equation}
which guarantees the compatibility of
Eq.~\eqref{eq:first-order-main} with the linearized Bianchi identity
\begin{equation}
    \nabla^{\mu}
    G_{\mu\nu}^{(1)}
    =
    0.
\end{equation}

A useful consistency check is provided by considering a linear function in \(Y\),
\begin{equation}
    \Psi
    =
    \alpha Y,
\end{equation}
where \(\alpha\) is constant. In this case,
\begin{equation}
    \Psi_R^{(0)}=0,
    \qquad
    \Psi_Y^{(0)}=\alpha,
    \qquad
    \nabla_{\mu}\Psi_Y^{(0)}=0.
\end{equation}
Equation~\eqref{eq:first-order-main} reduces to
\begin{equation}
    G_{\mu\nu}^{(1)}
    =
    \alpha
    \left[
    g_{\mu\nu}^{(0)}Y^{(0)}
    -
    4R_{\mu\alpha\beta\gamma}^{(0)}
    R_{\nu}^{(0)\,\alpha\beta\gamma}
    \right].
    \label{eq:linear-Y-source}
\end{equation}
In four dimensions, a Ricci-flat geometry satisfies
\begin{equation}
    4R_{\mu\alpha\beta\gamma}^{(0)}
    R_{\nu}^{(0)\,\alpha\beta\gamma}
    =
    g_{\mu\nu}^{(0)}
    R_{\alpha\beta\gamma\delta}^{(0)}
    R^{(0)\,\alpha\beta\gamma\delta}
    =
    g_{\mu\nu}^{(0)}Y^{(0)}.
    \label{eq:four-dimensional-riemann-identity}
\end{equation}
The source therefore vanishes identically,
\begin{equation}
    G_{\mu\nu}^{(1)}
    =
    0.
\end{equation}
This is consistent with the fact that, on a four-dimensional Ricci-flat
background, a linear function of \(Y\) differs from the
Gauss--Bonnet density only by Ricci-based terms, which are perturbatively
rigid.

Nontrivial local sources arise instead from nonlinear dependence on the
Riemann-sector invariants. In that case,
\(\Psi^{(0)}\), \(\Psi_R^{(0)}\), or \(\Psi_Y^{(0)}\) may depend on the
nonvanishing background invariant \(Y^{(0)}\), and the derivative terms in
Eq.~\eqref{eq:first-order-main} need not vanish. This is the mechanism by
which the Riemann sector evades the Ricci-based rigidity theorem.

For completeness, we recall the form of the first-order Einstein tensor.
The perturbation of the connection is
\begin{equation}
    C^{\alpha}{}_{\beta\gamma}
    =
    \frac{1}{2}
    g^{(0)\alpha\rho}
    \left(
    \nabla_{\beta}h_{\gamma\rho}
    +
    \nabla_{\gamma}h_{\beta\rho}
    -
    \nabla_{\rho}h_{\beta\gamma}
    \right).
    \label{eq:connection-perturbation}
\end{equation}
The first-order Ricci tensor is
\begin{equation}
    R_{\mu\nu}^{(1)}
    =
    \nabla_{\alpha}
    C^{\alpha}{}_{\mu\nu}
    -
    \nabla_{\nu}
    C^{\alpha}{}_{\mu\alpha},
    \label{eq:ricci-perturbation}
\end{equation}
and, since the background is Ricci-flat,
\begin{equation}
    G_{\mu\nu}^{(1)}
    =
    R_{\mu\nu}^{(1)}
    -
    \frac{1}{2}
    g_{\mu\nu}^{(0)}
    g^{(0)\alpha\beta}
    R_{\alpha\beta}^{(1)}.
    \label{eq:einstein-perturbation}
\end{equation}

\subsection{Static and spherically symmetric perturbations}
\label{sec:sss-perturbations}

Although the perturbative construction and the Ricci-based rigidity theorem
do not require any spacetime symmetry, in the remainder of this work we focus
on static and spherically symmetric vacuum configurations. This restriction
allows us to obtain explicit black-hole solutions and to connect the metric
perturbations directly with strong-field observables.

The most general static, spherically symmetric first-order metric
perturbation can be written as
\begin{equation}
    h_{\mu\nu}dx^{\mu}dx^{\nu}
    =h_{tt}(r)\,dt^{2} +
    2h_{tr}(r)\,dt\,dr  +
    h_{rr}(r)\,dr^{2}+
    r^{2}h_{\Omega}(r)\,d\Omega^{2}.
    \label{eq:general-sss-perturbation}
\end{equation}
The mixed \(dt\,dr\) term can be removed by a first-order redefinition of
the time coordinate. The angular perturbation can be absorbed into a
first-order redefinition of the radial coordinate, so that \(r\) remains the
areal radius. We therefore work in a Schwarzschild-like areal-radius gauge,
for which
\begin{equation}
    h_{tr}(r)=0,
    \qquad
    h_{\Omega}(r)=0.
\end{equation}
The first-order perturbation then takes the form
\begin{equation}
    h_{\mu\nu}
    =
    \operatorname{diag}
    \left(
    h_{tt}(r),
    h_{rr}(r),
    0,
    0
    \right).
    \label{eq:metric-perturbation-ansatz}
\end{equation}

The line element is written as
\begin{equation}
    ds^{2}
    =
    H(r)\,dt^{2}
    +
    K(r)\,dr^{2}
    +
    r^{2}
    \left(
    d\theta^{2}
    +
    \sin^{2}\theta\,d\phi^{2}
    \right).
    \label{eq:metricFO1}
\end{equation}
To first order in \(\lambda\),
\begin{equation}
    H(r)
    =
    -F(r)
    +
    \lambda h_{tt}(r)
    +
    \mathcal O(\lambda^{2}),
    \label{eq:H-definition}
\end{equation}
and
\begin{equation}
    K(r)
    =
    F^{-1}(r)
    +
    \lambda h_{rr}(r)
    +
    \mathcal O(\lambda^{2}),
    \label{eq:K-definition}
\end{equation}
where
\begin{equation}
    F(r)
    =
    1-\frac{r_s}{r},
    \qquad
    r_s=2GM.
    \label{eq:F-definition}
\end{equation}
The coordinate \(r\) is therefore the areal radius, and \(r_s\) is the
Schwarzschild radius of the zeroth-order geometry.

Equations~\eqref{eq:metricFO1}--\eqref{eq:F-definition} define the metric
ansatz used in the subsequent analysis. The functions \(h_{tt}(r)\) and
\(h_{rr}(r)\) contain both the sourced higher-curvature response and possible
homogeneous solutions of the linearized Einstein equations. The latter must
be fixed by specifying the asymptotic mass, the normalization of the timelike
Killing vector, and the residual gauge conditions.

The main conclusion of this section can be summarized as a perturbative
selection rule. Curvature terms constructed solely from quantities
that vanish on a Ricci-flat background cannot source a new analytic vacuum
branch connected to GR. By contrast, nonlinear interactions involving the
full Riemann tensor can generate nonzero first-order sources because the
Kretschmann scalar is already nonvanishing at zeroth order. The power-law
Gauss--Bonnet model studied below provides an explicit realization of this
mechanism.

\section{Horizon structure and strong-field phenomenology}
\label{sec:phenomenology}

In the previous section, we established the perturbative framework and
specialized the metric deformation to static and spherically symmetric
vacuum configurations in Schwarzschild-like areal-radius gauge. We now
study the geometric, orbital, and thermodynamic consequences of the
general first-order perturbation introduced in
Eqs.~\eqref{eq:metricFO1}--\eqref{eq:F-definition}. The radial functions
\(h_{tt}(r)\) and \(h_{rr}(r)\) are kept arbitrary until the regularity
conditions at the horizon are imposed.

The geometric and geodesic results are model-independent within the
static and spherically symmetric case. They apply to any theory admitting
a vacuum solution that is analytic in the coupling parameter \(\lambda\),
continuously connected to Schwarzschild as \(\lambda\rightarrow0\), and
expressible through first order in the form of
Eqs.~\eqref{eq:metricFO1}--\eqref{eq:F-definition}. 

Unless stated otherwise, we assume an asymptotically flat exterior region
and normalize the timelike Killing coordinate so that
\begin{equation}
    H(r)\longrightarrow-1,
    \qquad
    K(r)\longrightarrow1,
    \qquad
    r\longrightarrow\infty.
    \label{eq:asymptotic-normalization-general}
\end{equation}

The outer Killing horizon and its area shift probe the near-horizon
geometry \cite{HawkingEllis1973,Wald:1984rg,Poisson2004}. Circular timelike
geodesics determine the marginally stable orbit, the radial epicyclic
frequency, and the periapsis advance in the small-eccentricity limit
\cite{Chandrasekhar1983,BardeenPressTeukolsky1972,
Ori:2000zn,Abramowicz:2011xu,Buonanno:1998gg,Bambi-book}, whereas unstable
circular null geodesics determine the photon sphere and the critical impact
parameter associated with the black-hole shadow \cite{Perlick:2021aok}.
These quantities depend on different combinations of \(h_{tt}\),
\(h_{rr}\), and their radial derivatives, and therefore provide a general
dictionary between a first-order metric deformation and its strong-field
consequences. The Wald entropy additionally retains explicit dependence on
derivatives of the higher-curvature Lagrangian evaluated on the background
horizon \cite{Wald1993,IyerWald1994}.

The results of this section do not yet constitute predictions of a specific
modified-gravity theory. Such predictions require solving the first-order
field equations and specifying both the metric perturbations and the
relevant Lagrangian derivatives. In Sec.~\ref{sec:power-law-gb}, the general relations derived below will
be specialized to \(\Psi_{\eta}\propto\mathcal G^{\eta}\), thereby
converting the model-independent map into explicit functions of
\(\eta\) and the effective dimensionless coupling \(\bar\lambda\)
introduced there.

Throughout this section, all expressions are consistently truncated at
first order in \(\lambda\). In the exterior region, the perturbative
description is valid only where the corrections remain small relative to
the corresponding Schwarzschild metric components. From
Eqs.~\eqref{eq:H-definition} and \eqref{eq:K-definition}, this requires
\begin{equation}
    \left|
    \lambda F(r)h_{rr}(r)
    \right|
    \ll 1,
    \qquad
    \left|
    \lambda F^{-1}(r)h_{tt}(r)
    \right|
    \ll 1.
    \label{eq:perturbative-validity}
\end{equation}
All orbital quantities derived below must be understood within this
domain of validity. For observables involving radial derivatives of the
metric, Eq.~\eqref{eq:perturbative-validity} must be supplemented by the
requirement that the corresponding derivative combinations remain
perturbatively small at the relevant orbital radius. In particular, the
first-order displacements of the horizon, ISCO, and photon sphere must
remain small compared with their Schwarzschild values.

The relative conditions in
Eq.~\eqref{eq:perturbative-validity} are not directly applicable at
\(r=r_s\), where Schwarzschild-like coordinates are singular. The horizon
displacement will instead be determined by expanding the functions that
define the horizon around their simple zeroth-order root and requiring
the Killing-horizon and null-surface conditions to identify the same
hypersurface.

\subsection{Killing horizon and first-order displacement}
\label{sec:horizon-shift}

We begin by determining the location of the outer horizon of the general
first-order geometry. We denote by \(r_H\) the
areal radius of the outer horizon of the perturbed spacetime. Thus,
\(r_H\rightarrow r_s\) in the general-relativistic limit
\(\lambda\rightarrow0\). Here \(M\) denotes the fixed asymptotic mass when
the standard ADM definition applies, and otherwise the mass parameter of
the reference Schwarzschild solution.

For the static and spherically symmetric line element introduced in
Eq.~\eqref{eq:metricFO1}, the inverse radial metric component is
\(g^{rr}=K^{-1}\). Because the expansion of \(K\) need not be uniform near
the zeroth-order horizon, it is convenient to regard the inverse radial
component as a primary perturbative quantity and write
\begin{equation}
    g^{rr}(r)
    =
    F(r)
    +
    \lambda F_1(r)
    +
    \mathcal O(\lambda^2),
    \label{eq:grr-F1-intro}
\end{equation}
where \(F(r)=1-r_s/r\) is the Schwarzschild metric function defined in
Eq.~\eqref{eq:F-definition}. For every fixed exterior point \(r>r_s\),
comparison with Eq.~\eqref{eq:K-definition} gives
\begin{equation}
    F_1(r)
    \equiv
    -F^2(r)h_{rr}(r).
    \label{eq:F1-definition-intro}
\end{equation}
The quantity \(F_1\) is the first-order correction to the inverse radial
metric component. Equation~\eqref{eq:F1-definition-intro} is to be
understood as a relation between perturbative coefficients in the exterior,
not as a uniform series inversion at \(F=0\). In the horizon analysis below,
we assume that \(F_1(r)\) admits a finite and continuous extension to
\(r=r_s\), even when \(h_{rr}\) itself has poles in Schwarzschild-like
coordinates.

The static Killing vector is
\begin{equation}
    \xi^\mu
    =
    \left(
    \frac{\partial}{\partial t}
    \right)^\mu,
    \label{eq:killing-vector}
\end{equation}
and its squared norm is
\begin{equation}
    \xi^\mu\xi_\mu
    =
    g_{\mu\nu}\xi^\mu\xi^\nu
    =
    H(r).
    \label{eq:killing-norm}
\end{equation}
The outer Killing horizon is therefore located at the largest root of
\begin{equation}
    H(r_H)=0.
    \label{eq:horizon-H-condition}
\end{equation}
For a regular static and asymptotically flat black-hole geometry, this
Killing horizon coincides with the event horizon under the usual global
assumptions \cite{HawkingEllis1973,Wald:1984rg,Poisson2004}.

We consider the horizon branch continuously connected to the Schwarzschild
solution and expand its radius as
\begin{equation}
    r_H
    =
    r_s
    +
    \lambda r_H^{(1)}
    +
    \mathcal O(\lambda^2),
    \label{eq:horizon-expansion}
\end{equation}
where \(r_H^{(1)}\) denotes the first-order displacement coefficient. The displacement of the horizon is therefore
\begin{equation}
    \delta r_H
    \equiv
    r_H-r_s
    =
    \lambda r_H^{(1)}
    +
    \mathcal O(\lambda^2).
    \label{eq:horizon-displacement-definition}
\end{equation}

Using
\begin{equation}
    H(r)
    =
    -F(r)
    +
    \lambda h_{tt}(r)
    +
    \mathcal O(\lambda^2),
\end{equation}
and assuming that \(h_{tt}(r)\) is finite and continuous at \(r=r_s\), the
Killing-horizon condition \eqref{eq:horizon-H-condition} gives
\begin{align}
    0
    &=
    -F(r_H)
    +
    \lambda h_{tt}(r_H)
    +
    \mathcal O(\lambda^2)
    \nonumber\\
    &=
    -\lambda r_H^{(1)}F'(r_s)
    +
    \lambda h_{tt}(r_s)
    +
    \mathcal O(\lambda^2).
    \label{eq:horizon-H-expanded}
\end{align}
Since
\begin{equation}
    F'(r_s)=\frac{1}{r_s},
    \label{eq:Fprime-horizon}
\end{equation}
we obtain
\begin{equation}
    r_H^{(1)}
    =
    r_s h_{tt}(r_s).
    \label{eq:horizon-shift-H}
\end{equation}
The horizon radius determined from the norm of the Killing vector is
therefore
\begin{equation}
    r_H
    =
    r_s
    +
    \lambda r_s h_{tt}(r_s)
    +
    \mathcal O(\lambda^2).
    \label{eq:horizon-radius-H}
\end{equation}

The same hypersurface must also be null. Because \(r\) is the areal-radius
coordinate, the normal to a constant-\(r\) hypersurface is
\(\nabla_\mu r\), whose squared norm is
\begin{equation}
    g^{\mu\nu}
    \nabla_\mu r
    \nabla_\nu r
    =
    g^{rr}.
    \label{eq:constant-r-normal}
\end{equation}
Consequently, a constant-\(r\) hypersurface is null when
\begin{equation}
    g^{rr}(r_H)=0.
    \label{eq:horizon-grr-condition}
\end{equation}

Substituting Eqs.~\eqref{eq:grr-F1-intro} and
\eqref{eq:horizon-expansion} into
Eq.~\eqref{eq:horizon-grr-condition}, and using the continuous extension of
\(F_1(r)\) to \(r=r_s\), we find
\begin{align}
    0
    &=
    F(r_H)
    +
    \lambda F_1(r_H)
    +
    \mathcal O(\lambda^2)
    \nonumber\\
    &=
    \lambda r_H^{(1)}F'(r_s)
    +
    \lambda F_1(r_s)
    +
    \mathcal O(\lambda^2).
    \label{eq:horizon-grr-expanded}
\end{align}
It follows that
\begin{equation}
    r_H^{(1)}
    =
    -r_s F_1(r_s).
    \label{eq:horizon-shift-grr}
\end{equation}

For a regular static black-hole solution, the surface on which the Killing
vector becomes null must coincide with the null constant-\(r\)
hypersurface. Comparing Eqs.~\eqref{eq:horizon-shift-H} and
\eqref{eq:horizon-shift-grr}, we obtain the consistency condition
\begin{equation}
    F_1(r_s)
    =
    -h_{tt}(r_s).
    \label{eq:horizon-consistency-F1}
\end{equation}
Using Eq.~\eqref{eq:F1-definition-intro}, this condition can equivalently
be written as
\begin{equation}
    \lim_{r\rightarrow r_s}
    F^2(r)h_{rr}(r)
    =
    h_{tt}(r_s).
    \label{eq:horizon-consistency-hrr}
\end{equation}

Equation~\eqref{eq:horizon-consistency-hrr} is not an identity for two
arbitrary and independent perturbations \(h_{tt}\) and \(h_{rr}\). It is a
necessary first-order condition for the perturbed solution to possess a
regular static horizon. In particular, \(h_{rr}\) itself need not remain
finite at \(r=r_s\) in Schwarzschild-like coordinates. The geometrically
relevant quantity entering the inverse metric is \(F_1=-F^2h_{rr}\), which
may remain finite even if \(h_{rr}\) contains poles at the zeroth-order
horizon. Such poles may therefore be coordinate effects rather than physical
curvature singularities.

The agreement between the Killing-horizon and null-surface conditions is a
necessary regularity check. Complete regularity additionally requires that
curvature invariants and the metric expressed in horizon-penetrating
coordinates remain finite at \(r=r_H\).

Combining the two equivalent horizon conditions, the general first-order
position of the outer horizon is
\begin{equation}
    r_H
    =
    r_s
    +
    \lambda r_s h_{tt}(r_s)
    +
    \mathcal O(\lambda^2)
    =
    r_s
    -
    \lambda r_s F_1(r_s)
    +
    \mathcal O(\lambda^2).
    \label{eq:horizon-radius-final}
\end{equation}
Hence, the surface \(r=r_s\) is not necessarily the horizon of the corrected
geometry. The leading displacement is determined either by the temporal
metric perturbation evaluated at the Schwarzschild horizon or,
equivalently, by the first-order correction to the inverse radial metric component.

Since \(r\) is the areal radius, the corrected horizon area is
\begin{equation}
    A_H
    =
    4\pi r_H^2
    =
    A_H^{(0)}
    +
    \lambda A_H^{(1)}
    +
    \mathcal O(\lambda^2),
    \label{eq:horizon-area-expansion}
\end{equation}
where
\begin{equation}
    A_H^{(0)}
    =
    4\pi r_s^2,
    \qquad
    A_H^{(1)}
    =
    8\pi r_s^2 h_{tt}(r_s).
    \label{eq:horizon-area-terms}
\end{equation}
This area displacement will provide the geometric contribution to the
first-order Wald entropy derived later in this section.

\subsection{Innermost stable circular orbit}
\label{sec:ISCO}

In Schwarzschild spacetime, the innermost stable circular orbit (ISCO) is the
smallest-radius stable circular timelike geodesic, and circular orbits below
it are radially unstable. A generic deformed geometry could in principle
contain additional disconnected bands of stable circular motion. Throughout
this work, we therefore define \(r_{\rm ISCO}\) as the marginally stable
circular-orbit branch continuously connected to the Schwarzschild value
\(3r_s\) as \(\lambda\rightarrow0\). This branch defines a characteristic
strong-field scale and plays an important role in accretion physics, the
transition from inspiral to plunge, and relativistic binary dynamics
\cite{Chandrasekhar1983,BardeenPressTeukolsky1972,
Ori:2000zn,Abramowicz:2011xu,Buonanno:1998gg,Bambi-book}.

To derive the ISCO condition, we first construct the radial effective
potential for geodesic motion in the static and spherically symmetric
geometry introduced in Eq.~\eqref{eq:metricFO1}. The geodesic Lagrangian is
\begin{equation}
    \mathcal{L}_{\rm geo}
    =
    \frac{1}{2}
    \left[
    H(r)\dot{t}^{\,2}
    +
    K(r)\dot{r}^{\,2}
    +
    r^{2}\dot{\theta}^{\,2}
    +
    r^{2}\sin^{2}\theta\,\dot{\phi}^{\,2}
    \right],
    \label{eq:geodesic-lagrangian}
\end{equation}
where a dot denotes differentiation with respect to an affine parameter
\(\chi\). For timelike geodesics, \(\chi\) may be chosen as proper time. By spherical symmetry, the motion can be restricted to the equatorial plane, $    \theta=\tfrac{\pi}{2}, \dot{\theta}=0.$

The reduced geodesic Lagrangian is then
\begin{equation}
    \mathcal{L}_{\rm geo}
    =
    \frac{1}{2}
    \left[
    H(r)\dot{t}^{\,2}
    +
    K(r)\dot{r}^{\,2}
    +
    r^{2}\dot{\phi}^{\,2}
    \right].
    \label{eq:equatorial-lagrangian}
\end{equation}

Since \(t\) and \(\phi\) are cyclic coordinates, their conjugate momenta
are conserved. We define
\begin{equation}
    H(r)\dot{t}
    =
    -E,
    \qquad
    r^{2}\dot{\phi}
    =
    L,
    \label{eq:conserved-quantities}
\end{equation}
where \(E\) and \(L\) are, respectively, the conserved energy and angular
momentum per unit rest mass for timelike geodesics.

The normalization of the tangent vector can be written as
\begin{equation}
    H(r)\dot{t}^{\,2}
    +
    K(r)\dot{r}^{\,2}
    +
    r^{2}\dot{\phi}^{\,2}
    =
    -k,
    \label{eq:geodesic-normalization}
\end{equation}
where
\begin{equation}
    k
    =
    \begin{cases}
        1, & \text{timelike geodesics},\\
        0, & \text{null geodesics}.
    \end{cases}
    \label{eq:k-geodesics}
\end{equation}
Substituting Eq.~\eqref{eq:conserved-quantities} into
Eq.~\eqref{eq:geodesic-normalization}, we obtain
\begin{equation}
    \frac{1}{2}
    \left(
    \frac{dr}{d\chi}
    \right)^2
    +
    V(r,E,L)
    =
    0,
    \label{eq:radial-effective-potential}
\end{equation}
where
\begin{equation}
    V(r,E,L)
    =
    \frac{E^2}{2H(r)K(r)}
    +
    \frac{1}{2K(r)}
    \left(
    k+\frac{L^2}{r^2}
    \right).
    \label{eq:effective-potential}
\end{equation}

Using Eqs.~\eqref{eq:H-definition} and \eqref{eq:K-definition},
\begin{equation}
    V
    =
    V_{\rm Schw}
    +
    \lambda\delta V
    +
    \mathcal O(\lambda^2),
    \label{eq:potential-expansion}
\end{equation}
where
\begin{equation}
    V_{\rm Schw}
    =
    -\frac{E^2}{2}
    +
    \frac{F(r)}{2}
    \left(
    k+\frac{L^2}{r^2}
    \right),
    \label{eq:schwarzschild-potential}
\end{equation}
and
\begin{equation}
     \delta V
    =\frac{E^2}{2}
    \left[
    Fh_{rr}
    -
    F^{-1}h_{tt}
    \right]-
    \frac{F^2h_{rr}}{2}
    \left(
    k+\frac{L^2}{r^2}
    \right).
    \label{eq:potential-correction}
\end{equation}

For timelike geodesics, \(k=1\). Circular orbits satisfy
\begin{equation}
    V(r_c,E,L)=0,
    \qquad
    \left.
    \frac{\partial V}{\partial r}
    \right|_{r=r_c}=0,
    \label{eq:circular-orbit-conditions}
\end{equation}
and stability requires
\begin{equation}
    \left.
    \frac{\partial^2V}{\partial r^2}
    \right|_{r=r_c}>0.
    \label{eq:circular-stability-condition}
\end{equation}
The ISCO is determined by 
\begin{equation}
    V(r_{\rm ISCO})=0,
    \qquad
    V'(r_{\rm ISCO})=0,
    \qquad
    V''(r_{\rm ISCO})=0.
    \label{eq:isco-conditions}
\end{equation}

Introduce
\begin{equation}
    \mathcal U(r;E,L)
    =
    \frac{E^2}{H(r)}
    +
    1
    +
    \frac{L^2}{r^2},
    \label{eq:radial-function-isco}
\end{equation}
so that
\begin{equation}
    V(r,E,L)
    =
    \frac{\mathcal U(r;E,L)}{2K(r)}.
    \label{eq:potential-factorized}
\end{equation}
At a circular orbit,
\begin{equation}
    \mathcal U(r_c;E,L)=0,
    \qquad
    \mathcal U'(r_c;E,L)=0.
    \label{eq:radial-circular-conditions}
\end{equation}
Consequently,
\begin{equation}
    V''(r_c,E,L)
    =
    \frac{\mathcal U''(r_c;E,L)}{2K(r_c)}.
    \label{eq:potential-second-derivative}
\end{equation}
Since \(K(r_c)>0\), marginal stability is equivalent to
\begin{equation}
    \mathcal U''(r_{\rm ISCO};E,L)=0.
    \label{eq:radial-marginal-condition}
\end{equation}

The circular-orbit conditions give
\begin{equation}
    E^2(r)
    =
    \frac{2H^2(r)}
    {rH'(r)-2H(r)},
    \qquad
    L^2(r)
    =
    \frac{r^3H'(r)}
    {2H(r)-rH'(r)}.
    \label{eq:EL-circular-general}
\end{equation}
The marginal-stability equation becomes
\begin{equation}
    H(r)H''(r)
    -
    2[H'(r)]^2
    +
    \frac{3}{r}H(r)H'(r)
    =
    0.
    \label{eq:isco-A-condition}
\end{equation}
Thus, in areal-radius gauge, the coordinate position of the ISCO depends
only on \(H(r)\).

Expanding
\begin{equation}
    r_{\rm ISCO}
    =
    3r_s
    +
    \lambda r_{\rm ISCO}^{(1)}
    +
    \mathcal O(\lambda^2),
    \label{eq:isco-expansion}
\end{equation}
we obtain
\begin{equation}
    r_{\rm ISCO}^{(1)}
    =
    3r_s
    \left[
    18r_s^2h_{tt}''(3r_s)
    +
    6r_sh_{tt}'(3r_s)
    +
    h_{tt}(3r_s)
    \right].
    \label{eq:isco-shift-first-order}
\end{equation}
Therefore,
\begin{align}
    r_{\rm ISCO}
    =
    {}&
    3r_s
    +
    3\lambda r_s
    \left[
    18r_s^2h_{tt}''(3r_s)
    +
    6r_sh_{tt}'(3r_s)
    +
    h_{tt}(3r_s)
    \right]+
    \mathcal O(\lambda^2).
    \label{eq:isco-radius}
\end{align}

\subsection{Radial epicyclic frequency and relativistic precession}
\label{sec:epicyclic-precession}

We now consider small radial perturbations around a stable circular timelike
geodesic. The resulting oscillations define the radial epicyclic frequency \cite{Berry2011}.
For a slightly eccentric orbit oscillating about the circular radius, its
difference from the azimuthal orbital frequency produces the relativistic
advance of the periapsis in the epicyclic, or small-eccentricity, limit.

Let
\begin{equation}
    r(\chi)=r_0+\delta r(\chi),
    \qquad
    |\delta r|\ll r_0.
    \label{eq:radial-perturbation}
\end{equation}
Linearizing the radial equation gives
\begin{equation}
    \frac{d^2\delta r}{d\chi^2}
    +
    \left.
    \frac{\partial^2V}{\partial r^2}
    \right|_{r=r_0}
    \delta r
    =
    0.
    \label{eq:radial-oscillation-affine}
\end{equation}
The coordinate-time radial frequency is
\begin{equation}
    \Omega_{\rm rad}^2
    =
    \frac{1}{\dot t^{\,2}}
    \left.
    \frac{\partial^2V}{\partial r^2}
    \right|_{r=r_0}.
    \label{eq:radial-frequency-definition}
\end{equation}
Using the circular-orbit conditions,
\begin{equation}
    \left.
    \frac{\partial^2V}{\partial r^2}
    \right|_{r=r_0}
    =
    \left.
    \frac{L^2}{K(r)r^3}
    \left[
    \frac{H''}{H'}
    -
    2\frac{H'}{H}
    +
    \frac{3}{r}
    \right]
    \right|_{r=r_0}.
    \label{eq:V-second-derivative}
\end{equation}

The azimuthal angular frequency is
\begin{equation}
    \Omega_\phi
    =
    \frac{d\phi}{dt},
    \qquad
    \Omega_\phi^2
    =
    -\frac{H'(r_0)}{2r_0}.
    \label{eq:orbital-angular-velocity-H}
\end{equation}
Thus,
\begin{equation}
    \Omega_\phi^2
    =
    \frac{r_s}{2r_0^3}
    -
    \frac{\lambda}{2r_0}h_{tt}'(r_0)
    +
    \mathcal O(\lambda^2).
    \label{eq:orbital-angular-velocity-expanded}
\end{equation}
The radial epicyclic frequency is
\begin{equation}
    \Omega_{\rm rad}^2
    =
    \frac{r_0\Omega_\phi^2}{K(r_0)}
    \left[
    \frac{H''(r_0)}{H'(r_0)}
    -
    2\frac{H'(r_0)}{H(r_0)}
    +
    \frac{3}{r_0}
    \right].
    \label{eq:radial-frequency-general}
\end{equation}
Expanding through first order,
\begin{equation}
    \Omega_{\rm rad}^2
    =
    \Omega_\phi^2
    \left[
    1-\frac{3r_s}{r_0}
    +
    \lambda\zeta(r_0,r_s)
    \right]
    +
    \mathcal O(\lambda^2),
    \label{eq:radial-frequency-expanded}
\end{equation}
where
\begin{align}
    \zeta(r_0,r_s)
    =
    {}&
    -\frac{r_0^2(r_0-r_s)}{r_s}h_{tt}''(r_0)
    +
    \frac{2r_0(2r_s-r_0)}{r_s}h_{tt}'(r_0)
    \nonumber\\
    &-
    \frac{2r_s}{r_0-r_s}h_{tt}(r_0)
    -
    \frac{(r_0-r_s)(r_0-3r_s)}{r_0^2}h_{rr}(r_0).
    \label{eq:zeta-definition}
\end{align}

As an internal consistency check, the factor multiplying \(h_{rr}\) vanishes
at \(r_0=3r_s\), and
\begin{equation}
    r_{\rm ISCO}^{(1)}
    =
    -3r_s\,\zeta(3r_s,r_s),
    \label{eq:isco-zeta-consistency}
\end{equation}
in exact agreement with Eq.~\eqref{eq:isco-shift-first-order}.

In Schwarzschild,
\begin{equation}
    \Omega_{\phi,\rm Schw}^2
    =
    \frac{r_s}{2r_0^3},
    \qquad
    \Omega_{\rm rad,Schw}^2
    =
    \Omega_{\phi,\rm Schw}^2
    \left(
    1-\frac{3r_s}{r_0}
    \right).
    \label{eq:schwarzschild-epicyclic-frequency}
\end{equation}

For a slightly eccentric orbit about \(r_0\), the periapsis advance is
\begin{equation}
    \vartheta
    =
    2\pi
    \left(
    \frac{\Omega_\phi}{\Omega_{\rm rad}}-1
    \right).
    \label{eq:periapsis-definition}
\end{equation}
For Schwarzschild,
\begin{equation}
    \vartheta_{\rm Schw}
    =
    2\pi
    \left[
    \left(
    1-\frac{3r_s}{r_0}
    \right)^{-1/2}
    -
    1
    \right].
    \label{eq:schwarzschild-periapsis}
\end{equation}
In the weak-field regime,
\begin{equation}
    \vartheta_{\rm Schw}
    =
    \frac{3\pi r_s}{r_0}
    +
    \mathcal O\left(\frac{r_s^2}{r_0^2}\right)
    =
    \frac{6\pi GM}{r_0}
    +
    \mathcal O\left(\frac{G^2M^2}{r_0^2}\right).
\end{equation}
The first-order expansion requires
\begin{equation}
    |\lambda\zeta(r_0,r_s)|
    \ll
    1-\frac{3r_s}{r_0},
    \label{eq:precession-validity}
\end{equation}
and gives
\begin{equation}
    \Delta\vartheta
    =
    -\lambda\pi\zeta(r_0,r_s)
    \left(
    1-\frac{3r_s}{r_0}
    \right)^{-3/2}
    +
    \mathcal O(\lambda^2).
    \label{eq:periapsis-correction}
\end{equation}
The divergence as \(r_0\rightarrow3r_s\) signals the breakdown of the
expansion around the Schwarzschild radial frequency, rather than a physical
singularity.

\subsection{Photon sphere and black-hole shadow}
\label{sec:photon-shadow}

We now consider null geodesics and the shadow associated with
the perturbed black-hole spacetime. We use the asymptotic normalization
fixed in Eq.~\eqref{eq:asymptotic-normalization-general}, which ensures that the conserved energy \(E\), and hence the impact parameter \(b=L/E\), have their standard asymptotic interpretation \cite{Perlick:2021aok}.

For null geodesics,
\begin{equation}
    V_{\rm null}
    =
    \frac{1}{2K(r)}
    \left[
    \frac{E^2}{H(r)}
    +
    \frac{L^2}{r^2}
    \right].
    \label{eq:null-effective-potential}
\end{equation}
A circular null orbit satisfies
\begin{equation}
    V_{\rm null}(r_{\rm ph})=0,
    \qquad
    V_{\rm null}'(r_{\rm ph})=0.
    \label{eq:null-circular-conditions}
\end{equation}
Defining \(b=L/E\),
\begin{equation}
    b^2(r)
    =
    -\frac{r^2}{H(r)}.
    \label{eq:impact-parameter-function}
\end{equation}
The photon-sphere condition is
\begin{equation}
    H'(r_{\rm ph})
    -
    \frac{2H(r_{\rm ph})}{r_{\rm ph}}
    =
    0.
    \label{eq:photon-sphere-condition}
\end{equation}
Thus, \(r_{\rm ph}\) is independent of \(K(r)\).

Expanding around
\begin{equation}
    r_{\rm ph}^{(0)}
    =
    \frac{3}{2}r_s,
\end{equation}
we obtain
\begin{equation}
    r_{\rm ph}^{(1)}
    =
    -\frac{3r_s}{8}
    \left[
    3r_s h_{tt}'\!\left(\frac{3r_s}{2}\right)
    -
    4h_{tt}\!\left(\frac{3r_s}{2}\right)
    \right],
    \label{eq:photon-sphere-shift-first-order}
\end{equation}
and therefore
\begin{align}
    r_{\rm ph}
    =
    {}&
    \frac{3}{2}r_s
    -
    \frac{3\lambda r_s}{8}
    \left[
    3r_s h_{tt}'\!\left(\frac{3r_s}{2}\right)
    -
    4h_{tt}\!\left(\frac{3r_s}{2}\right)
    \right]
    \nonumber\\
    &+
    \mathcal O(\lambda^2).
    \label{eq:photon-sphere-radius}
\end{align}

The relevant orbit must be unstable,
\begin{equation}
    V_{\rm null}''(r_{\rm ph})<0.
\end{equation}
The critical impact parameter is
\begin{equation}
    b_c
    =
    \frac{r_{\rm ph}}
    {\sqrt{-H(r_{\rm ph})}}.
    \label{eq:critical-impact-parameter}
\end{equation}
We denote the critical shadow impact parameter for an asymptotic observer by
\begin{equation}
    b_{\rm sh}\equiv b_c.
    \label{eq:shadow-radius-definition}
\end{equation}
For a static observer at \(r_{\rm obs}\),
\begin{equation}
    \sin\alpha_{\rm sh}
    =
    \frac{b_c}{r_{\rm obs}}
    \sqrt{-H(r_{\rm obs})}.
    \label{eq:shadow-angular-radius}
\end{equation}

For Schwarzschild,
\begin{equation}
    b_{\rm sh}^{(0)}
    =
    \frac{3\sqrt3}{2}r_s.
\end{equation}
Because the background impact-parameter function is stationary at the
Schwarzschild photon sphere, \(r_{\rm ph}^{(1)}\) does not contribute
explicitly to the first-order shadow shift. One obtains
\begin{equation}
    b_{\rm sh}^{(1)}
    =
    \frac{9\sqrt3}{4}r_s
    h_{tt}\!\left(\frac{3r_s}{2}\right),
    \label{eq:shadow-shift-first-order}
\end{equation}
and
\begin{equation}
    b_{\rm sh}
    =
    \frac{3\sqrt3}{2}r_s
    +
    \frac{9\sqrt3}{4}
    \lambda r_s
    h_{tt}\!\left(\frac{3r_s}{2}\right)
    +
    \mathcal O(\lambda^2).
    \label{eq:shadow-radius}
\end{equation}

The photon-sphere radius and critical shadow impact parameter are both independent of
\(h_{rr}\) in areal-radius gauge, but probe different information:
\(r_{\rm ph}\) depends on \(h_{tt}\) and \(h_{tt}'\), whereas the leading
shadow correction depends only on \(h_{tt}\) evaluated at the unperturbed
photon sphere.

\subsection{First-order Wald entropy correction}
\label{sec:wald-entropy-general}

With
\begin{equation}
    \mathcal L_{\rm grav}
    =
    \frac{R}{2}
    +
    \lambda\Psi(R,X,Y),
\end{equation}
the Wald entropy \cite{Wald:1999vt} is
\begin{equation}
    S_{\rm W}
    =
    -\frac{2\pi}{\kappa^2}
    \int_{\mathcal B}
    d^2x\,\sqrt{\gamma}\,
    \frac{\partial\mathcal L_{\rm grav}}
    {\partial R_{\mu\nu\rho\sigma}}
    \epsilon_{\mu\nu}\epsilon_{\rho\sigma}.
    \label{eq:wald-formula}
\end{equation}
We assume that \(\Psi_R\), \(\Psi_X\), and \(\Psi_Y\) are finite and
continuous at
\begin{equation}
    (R,X,Y)
    =
    \left(
    0,0,\frac{12}{r_s^4}
    \right).
    \label{eq:wald-regularity-assumption}
\end{equation}

The curvature derivative is
\begin{align}
    \frac{\partial\mathcal L_{\rm grav}}
    {\partial R_{\mu\nu\rho\sigma}}
    =
    {}&
    \left(
    \frac12+\lambda\Psi_R
    \right)
    g^{\mu[\rho}g^{\sigma]\nu}+
    2\lambda\Psi_YR^{\mu\nu\rho\sigma}
    \nonumber\\
    &+
    \frac{\lambda\Psi_X}{2}
    \left(
    g^{\mu\rho}R^{\nu\sigma}
    -
    g^{\mu\sigma}R^{\nu\rho}
    -
    g^{\nu\rho}R^{\mu\sigma}
    +
    g^{\nu\sigma}R^{\mu\rho}
    \right)
    \label{eq:wald-riemann-derivative}
\end{align}
Hence,
\begin{equation}
    S_{\rm W}
    =
    \frac{4\pi}{\kappa^2}
    \int_{\mathcal B}
    d^2x\,\sqrt{\gamma}
    \left[
    \frac12
    +
    \lambda\Psi_R
    +
    \lambda\Psi_XR_{\mu\nu}g_\perp^{\mu\nu}
    -
    \lambda\Psi_YR^{\mu\nu\rho\sigma}
    \epsilon_{\mu\nu}\epsilon_{\rho\sigma}
    \right].
    \label{eq:wald-general-integrand}
\end{equation}

On Schwarzschild,
\begin{equation}
    R^{(0)}=0,
    \qquad
    R_{\mu\nu}^{(0)}=0,
    \qquad
    X^{(0)}=0,
\end{equation}
so the explicit \(\Psi_X\) term vanishes at first order. Moreover,
\begin{equation}
    Y_H^{(0)}
    =
    \frac{12}{r_s^4},
\end{equation}
and
\begin{equation}
    \left.
    R^{(0)\mu\nu\rho\sigma}
    \epsilon_{\mu\nu}^{(0)}
    \epsilon_{\rho\sigma}^{(0)}
    \right|_{r=r_s}
    =
    -\frac{4}{r_s^2}.
\end{equation}

Defining
\begin{equation}
    \Psi_A^H
    =
    \left.
    \Psi_A(R,X,Y)
    \right|_{R=0,\,X=0,\,Y=12/r_s^4},
    \qquad
    A=R,X,Y,
\end{equation}
the entropy through first order becomes
\begin{equation}\label{eq:wald-first-order-area}
    S_{\rm W}
    =
    \frac{A_H^{(0)}}{4G}
    +
    \frac{\lambda}{4G}
    \left\{
    A_H^{(1)}
    +
    2A_H^{(0)}
    \left[
    \Psi_R^H
    +
    \frac{4}{r_s^2}\Psi_Y^H
    \right]
    \right\}
    +
    \mathcal O(\lambda^2).
\end{equation}
Using
\begin{equation}
    A_H^{(1)}
    =
    2A_H^{(0)}h_{tt}(r_s),
\end{equation}
we finally obtain
\begin{equation}
    S_{\rm W}
    =
    \frac{A_H^{(0)}}{4G}
    +
    \frac{\lambda A_H^{(0)}}{2G}
    \left[
    h_{tt}(r_s)
    +
    \Psi_R^H
    +
    \frac{4}{r_s^2}\Psi_Y^H
    \right]
    +
    \mathcal O(\lambda^2).
    \label{eq:wald-first-order-final}
\end{equation}

The term \(h_{tt}(r_s)\) arises from the displacement of the horizon area,
whereas the terms involving \(\Psi_R^H\) and \(\Psi_Y^H\) are explicit
Noether-charge contributions. For a linear Gauss--Bonnet term, the local
metric perturbation vanishes but the entropy retains an additive topological
constant. For fixed spherical horizon topology, this constant does not
affect the entropy variations entering the first law, although it remains
present in the Wald Noether charge \cite{Wald1993}.

\section{Power-law Gauss--Bonnet gravity at first order}
\label{sec:power-law-gb}

We now apply the first-order framework developed in the previous sections
to a nonlinear power of the Gauss--Bonnet invariant. On a Ricci-flat
background, the Ricci scalar and Ricci tensor vanish, whereas the Riemann
tensor and the Kretschmann scalar remain nonzero. Consequently,
Riemann-dependent contributions can source nontrivial perturbations of the
Schwarzschild geometry even when purely Ricci-based analytic corrections
do not.

The Gauss--Bonnet invariant,
\(\mathcal G=R^2-4R_{\mu\nu}R^{\mu\nu}
+R_{\mu\nu\rho\sigma}R^{\mu\nu\rho\sigma}\), is the quadratic Lovelock
density. In four dimensions, a term linear in \(\mathcal G\), with a
constant coefficient, is topological and does not modify the local metric
field equations. Nonlinear functions of \(\mathcal G\), however, produce
genuine local contributions \cite{Lanczos1938,Lovelock,
NojiriOdintsov2005,DeFeliceTsujikawaCosmology2009}.

We consider
\begin{equation}
    \Psi_\eta(R,X,Y)
    =
    \ell_\lambda^{-2+4\eta}\mathcal G^\eta,
    \qquad
    \eta>0,
    \label{eq:power-law-gb-model}
\end{equation}
where
\begin{equation}
    \mathcal G=R^2-4X+Y,
    \qquad
    X\equiv R_{\mu\nu}R^{\mu\nu},
    \qquad
    Y\equiv
    R_{\mu\nu\rho\sigma}R^{\mu\nu\rho\sigma}.
    \label{eq:gb-invariant}
\end{equation}
The gravitational Lagrangian density is
\begin{equation}
    \mathcal L_{\rm grav}
    =
    \frac{R}{2}
    +
    \lambda\Psi_\eta(R,X,Y).
    \label{eq:power-law-gb-lagrangian}
\end{equation}
The length scale \(\ell_\lambda\) ensures that \(\Psi_\eta\) has the
same mass dimension as \(R\).

For noninteger \(\eta\), the function \(\mathcal G^\eta\) is not analytic
at \(\mathcal G=0\). Such models should therefore not be interpreted as
individual terms in an analytic curvature expansion about Minkowski
spacetime. In the present analysis, the perturbative expansion is in the
coupling \(\lambda\), and the real branch is evaluated on the exterior
Schwarzschild geometry, where \(\mathcal G^{(0)}>0\). The noninteger
power-law family is consequently regarded as a phenomenological
parametrization of nonlinear curvature effects rather than as a unique
ultraviolet completion.

On the Schwarzschild background,
\begin{equation}
    R^{(0)}=0,
    \qquad
    R_{\mu\nu}^{(0)}=0,
    \qquad
    X^{(0)}=0,
\end{equation}
and the Gauss--Bonnet invariant reduces to the Kretschmann scalar,
\begin{equation}
    \mathcal G^{(0)}
    =
    Y^{(0)}
    =
    \frac{12r_s^2}{r^6}.
    \label{eq:gb-background-schwarzschild}
\end{equation}
It therefore provides a radially varying source for the first-order
metric perturbation.

We introduce the dimensionless coupling
\begin{equation}
    \bar\lambda
    \equiv
    \lambda
    \left(
        \frac{\ell_\lambda}{r_s}
    \right)^{4\eta-2}.
    \label{eq:lambdabar-definition-power-gb}
\end{equation}
Dimensionless observables constrain \(\bar\lambda\), rather than
\(\lambda\) and \(\ell_\lambda\) separately. The explicit dependence on
\(r_s\) also implies that black holes with different masses probe the
same fundamental length scale with different effective strengths.

\subsection{First-order solution to the field equations}
\label{sec:power-law-gb-equations}

The quantities entering the effective first-order source are
\begin{equation}
    \Psi_\eta^{(0)}
    =
    \ell_\lambda^{-2+4\eta}
    \left(\mathcal G^{(0)}\right)^\eta,
    \qquad
    \Psi_{\eta,\mathcal G}^{(0)}
    =
    \eta\ell_\lambda^{-2+4\eta}
    \left(\mathcal G^{(0)}\right)^{\eta-1}.
\end{equation}
Since \(\Psi_\eta\) depends on \(R,X,Y\) only through \(\mathcal G\),
\begin{equation}
    \Psi_{\eta,R}
    =
    2R\Psi_{\eta,\mathcal G},
    \qquad
    \Psi_{\eta,X}
    =
    -4\Psi_{\eta,\mathcal G},
    \qquad
    \Psi_{\eta,Y}
    =
    \Psi_{\eta,\mathcal G}.
\end{equation}
Consequently,
\begin{equation}
    \Psi_{\eta,R}^{(0)}=0,
    \qquad
    \Psi_{\eta,Y}^{(0)}
    =
    \Psi_{\eta,\mathcal G}^{(0)}.
\end{equation}

Using the four-dimensional Ricci-flat identity
\begin{equation}
    4R_{\alpha\beta\gamma\mu}^{(0)}
    {R^{(0)\alpha\beta\gamma}}_{\nu}
    =
    g_{\mu\nu}^{(0)}\mathcal G^{(0)},
\end{equation}
the first-order vacuum equations become
\begin{align}
    G_{\mu\nu}^{(1)} =    -8\eta\ell_\lambda^{-2+4\eta}
    {R_{\mu\rho\nu}^{(0)}}{}^\sigma
    \nabla^\rho\nabla_\sigma
    \left[\left(\mathcal G^{(0)}\right)^{\eta-1}
    \right]- (\eta-1)\ell_\lambda^{-2+4\eta}
    \left(\mathcal G^{(0)}\right)^\eta
    g_{\mu\nu}^{(0)}.
    \label{eq:first-order-gb-source-explicit}
\end{align}
At \(\eta=1\), the derivative in the first term acts on a constant and
the second term vanishes. This is the local manifestation of the
topological character of the linear four-dimensional Gauss--Bonnet
density.

In the static and spherically symmetric areal-radius gauge, we write
\begin{equation}
    h_{\mu\nu}
    =
    \operatorname{diag}
    \left(
        h_{tt}(r),
        h_{rr}(r),
        0,
        0
    \right).
    \label{eq:power-gb-perturbation-ansatz}
\end{equation}
For the generic case \(\eta\neq1/2\), direct integration of the
\(tt\) and \(rr\) components of
Eq.~\eqref{eq:first-order-gb-source-explicit} gives 
\begin{equation}
    h_{tt}(r)
    =
    A(\eta)
    \frac{(6\eta-1)r_s-8\eta r}{r^{6\eta-2}}
    +
    \frac{c_1}{r}
    +
    c_2\left(1-\frac{r_s}{r}\right),
    \label{eq:htt-power-gb-general}
\end{equation}
and
\begin{equation}
    h_{rr}(r)
    =
    -A(\eta)
    \frac{
        12\eta(2\eta-1)r
        +2\eta r_s(7-12\eta)
        +r_s
    }{
        r^{6\eta-4}(r-r_s)^2
    }
    +
    \frac{c_1r}{(r-r_s)^2},
    \label{eq:hrr-power-gb-general}
\end{equation}
where
\begin{equation}
    A(\eta)
    =
    \frac{
        3^{\eta-1}4^\eta
        \ell_\lambda^{-2+4\eta}
        (\eta-1)r_s^{2\eta-1}
    }{
        2\eta-1
    }.
    \label{eq:Aeta-definition}
\end{equation}

The constant \(c_2\) changes the normalization of the timelike Killing
vector and is removed by imposing \(g_{tt}\rightarrow-1\) at infinity.
The integration constant \(c_1\) controls a source-free  \(1/r\) contribution corresponding to a shift of the Schwarzschild mass
\begin{equation}
    r_s\longrightarrow r_s+\lambda c_1.
\end{equation}
We select the asymptotically flat branch containing no additional contributions that shift the mass by imposing
\begin{equation}
    c_1=0,
    \qquad
    c_2=0.
    \label{eq:power-gb-integration-constants}
\end{equation}
As discussed below, the interpretation of \(c_1=0\) as a fixed-ADM-mass
condition is valid without ambiguity only for \(\eta>2/3\).

The sourced perturbations are
\begin{equation}
    h_{tt}(r)
    =
    A(\eta)
    \frac{(6\eta-1)r_s-8\eta r}{r^{6\eta-2}},
    \label{eq:htt-power-gb}
\end{equation}
and
\begin{equation}
    h_{rr}(r)
    =
    -\frac{A(\eta)}{r^{6\eta-4}}
    \left[
        \frac{12\eta(2\eta-1)}{r-r_s}
        +
        \frac{(2\eta+1)r_s}{(r-r_s)^2}
    \right].
    \label{eq:hrr-power-gb}
\end{equation}

The factor \(\eta-1\) in \(A(\eta)\) implies
\begin{equation}
    h_{tt}=h_{rr}=0,
    \qquad
    \eta=1.
\end{equation}
This conclusion is stronger than a first-order cancellation. At
\(\eta=1\), the complete theory differs from Einstein gravity only by a
topological term, and hence the local Schwarzschild geometry and all
local geodesic observables remain unchanged at every order in
\(\lambda\).

The value \(\eta=1/2\) does not belong to the generic branch because
\(A(\eta)\) contains the pole \((2\eta-1)^{-1}\). The field equations must
therefore be integrated independently at this value. Exact static
solutions of models containing \(R+\sqrt{\mathcal G}\) are known
\cite{MyrzakulovSebastianiZerbini2013}, but they need not lie on the same
asymptotically Schwarzschild solution studied here.

For generic \(\eta\neq1/2\), the leading large-radius behavior is
\begin{equation}
    h_{tt}(r)
    \sim
    -8\eta A(\eta)r^{3-6\eta},
    \qquad
    h_{rr}(r)
    \sim
    -12\eta(2\eta-1)A(\eta)r^{3-6\eta}.
    \label{eq:power-gb-asymptotic-behavior}
\end{equation}
This separates several physically distinct domains:

\begin{itemize}

\item When \(0<\eta<1/2\), the perturbations grow at infinity. Hence the
metric is not asymptotically Schwarzschild, and the expansion eventually
fails for every fixed nonzero coupling.

\item For \(\eta>1/2\), the perturbations tend to zero, so the metric
approaches Schwarzschild in the weak sense \(g_{\mu\nu}
-g_{\mu\nu}^{\rm Schw}\rightarrow0\).

\item For \(1/2<\eta<2/3\), the perturbations vanish at large
radius but decay more slowly than the Schwarzschild \(1/r\) mass term.
Thus, although the metric components approach their Schwarzschild values,
the spacetime does not satisfy the usual asymptotic conditions required
for the standard ADM mass to be well defined.

\item 
At \(\eta=2/3\), the sourced perturbation contains a \(1/r\) term with
the same radial dependence as the Schwarzschild mass term. Both this
sourced term and the source-free \(1/r\) contribution proportional to
\(c_1\) modify the coefficient that determines the ADM mass.
Consequently, setting \(c_1=0\) removes the source-free contribution but
does not, by itself, keep the physical ADM mass fixed.

\item For \(\eta>2/3\), the sourced perturbation decays faster
than \(1/r\) and therefore does not modify the asymptotic mass term. The
only additional \(1/r\) contribution is the homogeneous mode proportional
to \(c_1\). Setting \(c_1=0\) then selects the branch with the same ADM
mass as the reference Schwarzschild solution.

\end{itemize}

We therefore use
\begin{equation}
    \eta>\frac{2}{3}
    \label{eq:eta-fixed-adm-domain}
\end{equation}
as the strict domain for the fixed-ADM phenomenological analysis.
Expressions displayed for \(1/2<\eta\leq2/3\) should be understood only
as the formal continuation of the minimally sourced solution.

Finally, perturbative control requires the dimensionless corrections to
remain much smaller than unity. Since the coefficients contain
\((2\eta-1)^{-1}\), values close to \(\eta=1/2\) require a
correspondingly smaller \(\bar\lambda\); the divergence of the
first-order coefficients signals loss of perturbative control, not a
physical divergence of an observable.

\subsection{Horizon displacement and coordinate-consistency check}
\label{sec:horizon-power-law-gb}

The outer horizon continuously connected to Schwarzschild satisfies
\begin{equation}
    r_H
    =
    r_s+\lambda r_sh_{tt}(r_s)
    +\mathcal O(\lambda^2).
\end{equation}
From Eq.~\eqref{eq:htt-power-gb},
\begin{equation}
    h_{tt}(r_s)
    =
    -(2\eta+1)A(\eta)r_s^{3-6\eta}.
\end{equation}
The corrected horizon radius is therefore
\begin{equation}
    \frac{r_H}{r_s}
    =
    1-\bar\lambda\mathcal C_H(\eta)
    +\mathcal O(\bar\lambda^2),
    \label{eq:horizon-radius-power-gb}
\end{equation}
where
\begin{equation}
    \mathcal C_H(\eta)
    =
    \frac{
        3^{\eta-1}4^\eta
        (2\eta+1)(\eta-1)
    }{
        2\eta-1
    }.
    \label{eq:horizon-coefficient-power-gb}
\end{equation}
Thus,
\begin{equation}
    \frac{\delta r_H}{r_s}
    =
    -\bar\lambda\mathcal C_H(\eta)
    +\mathcal O(\bar\lambda^2).
\end{equation}

For \(\bar\lambda>0\), the formal solution predicts an outward
displacement for \(1/2<\eta<1\) and an inward displacement for
\(\eta>1\). For \(\bar\lambda<0\), these directions are reversed. At
\(\eta=1\), the Schwarzschild horizon is unchanged exactly, rather than
only through first order.

Perturbative control requires
\begin{equation}
    \left|
        \bar\lambda\mathcal C_H(\eta)
    \right|
    \ll1.
    \label{eq:horizon-validity-power-gb}
\end{equation}
Large displacements near \(\eta=1/2\) indicate the failure of this
condition and must not be interpreted as quantitative predictions.

\begin{figure}[h]
    \centering
    \includegraphics[width=0.7\textwidth]
    {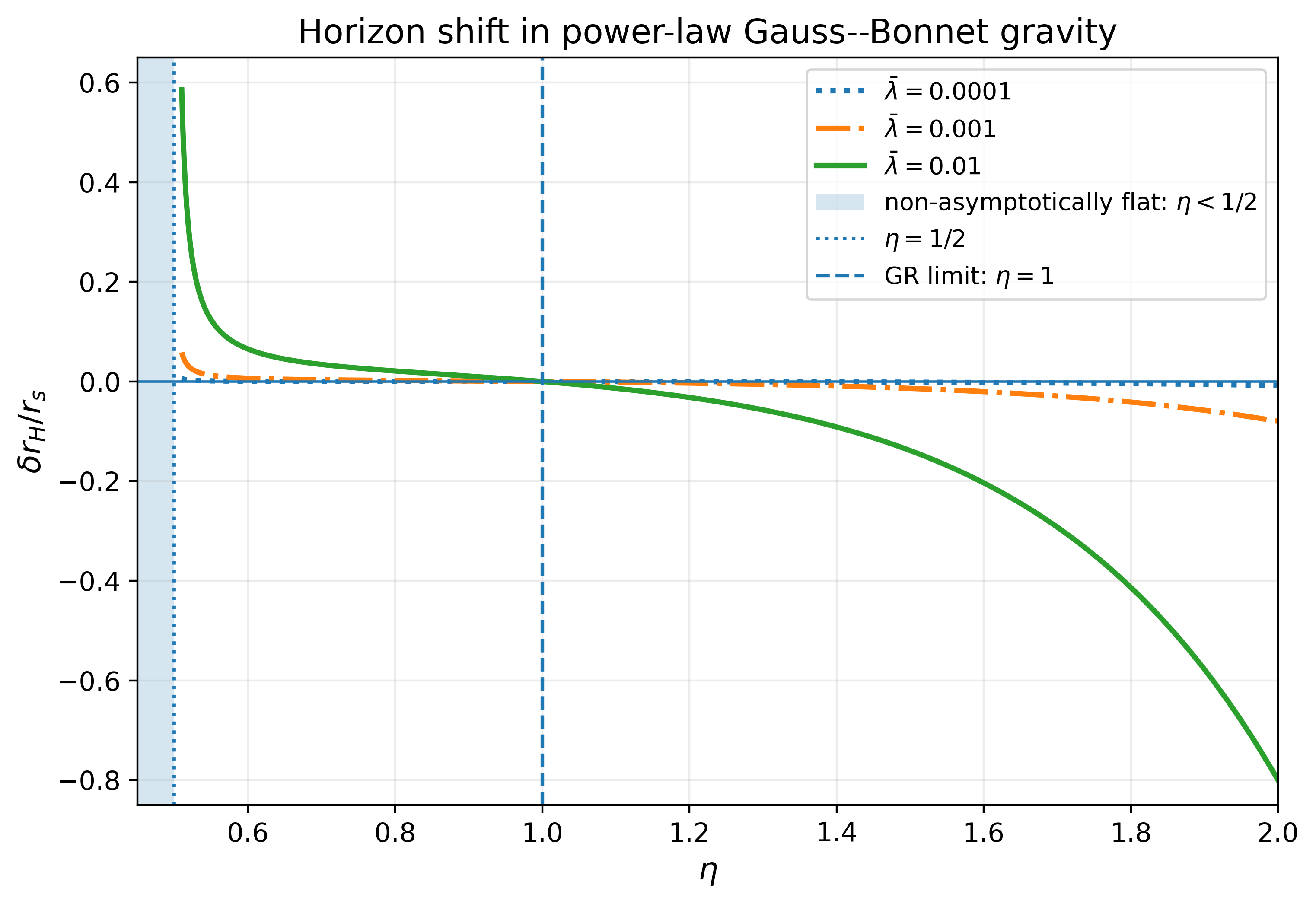}
    \caption{First-order fractional horizon displacement for
    representative positive values of \(\bar\lambda\). The divergence
    near \(\eta=1/2\) marks the breakdown of the generic perturbative
    branch. The interval \(1/2<\eta\leq2/3\), when shown, represents only
    the formal minimally sourced solution and is not used in the
    fixed-ADM phenomenological analysis. Portions for which
    \(|\bar\lambda\mathcal C_H|\) is not small are outside perturbative
    control.}
    \label{fig:horizon-shift-power-gb}
\end{figure}

Figure~\ref{fig:horizon-shift-power-gb} displays the formal horizon
displacement over a wider range of \(\eta\). Any displayed segment that
violates Eq.~\eqref{eq:horizon-validity-power-gb}, whether near
\(\eta=1/2\) or at larger \(\eta\), is illustrative only and is not a
quantitative first-order prediction.

To analyze the radial component, define
\begin{equation}
    \mathcal A(r)
    \equiv
    -g_{tt}(r)
    =
    F(r)-\lambda h_{tt}(r),
\end{equation}
and
\begin{equation}
    \mathcal B^{-1}(r)
    \equiv
    g^{rr}(r)
    =
    F(r)+\lambda F_1(r),
    \qquad
    F_1(r)\equiv-F^2(r)h_{rr}(r).
\end{equation}
Using Eq.~\eqref{eq:hrr-power-gb}, one obtains the finite function
\begin{equation}
    F_1(r)
    =
    \frac{A(\eta)}{r^{6\eta-2}}
    \left[
        12\eta(2\eta-1)(r-r_s)
        +(2\eta+1)r_s
    \right].
    \label{eq:F1-power-gb}
\end{equation}
At \(r=r_s\),
\begin{equation}
    F_1(r_s)
    =
    (2\eta+1)A(\eta)r_s^{3-6\eta}
    =
    -h_{tt}(r_s).
    \label{eq:horizon-consistency-power-gb}
\end{equation}
Consequently, the zero of the Killing norm,
\(\mathcal A(r_H)=0\), coincides through first order with the null
constant-\(r\) hypersurface defined by
\(\mathcal B^{-1}(r_H)=0\).

The poles in \(h_{rr}\) at \(r=r_s\) arise from expanding the singular
Schwarzschild component \(g_{rr}\) about the unperturbed horizon. They
disappear from \(F_1=-F^2h_{rr}\). Formally introducing an ingoing
coordinate
\begin{equation}
    dv
    =
    dt+
    \frac{dr}{
        \sqrt{\mathcal A(r)\mathcal B^{-1}(r)}
    },
\end{equation}

the radial-temporal metric becomes
\begin{equation}
    ds^2
    =
    -\mathcal A\,dv^2
    +
    2\sqrt{
        \frac{\mathcal A}{\mathcal B^{-1}}
    }\,dv\,dr
    +
    r^2d\Omega^2.
\end{equation}
The common simple zero makes the ratio
\(\mathcal A/\mathcal B^{-1}\) finite at the corrected horizon. This is
a necessary coordinate-consistency check.

\begin{figure}[h]
    \centering
    \includegraphics[width=\textwidth]
    {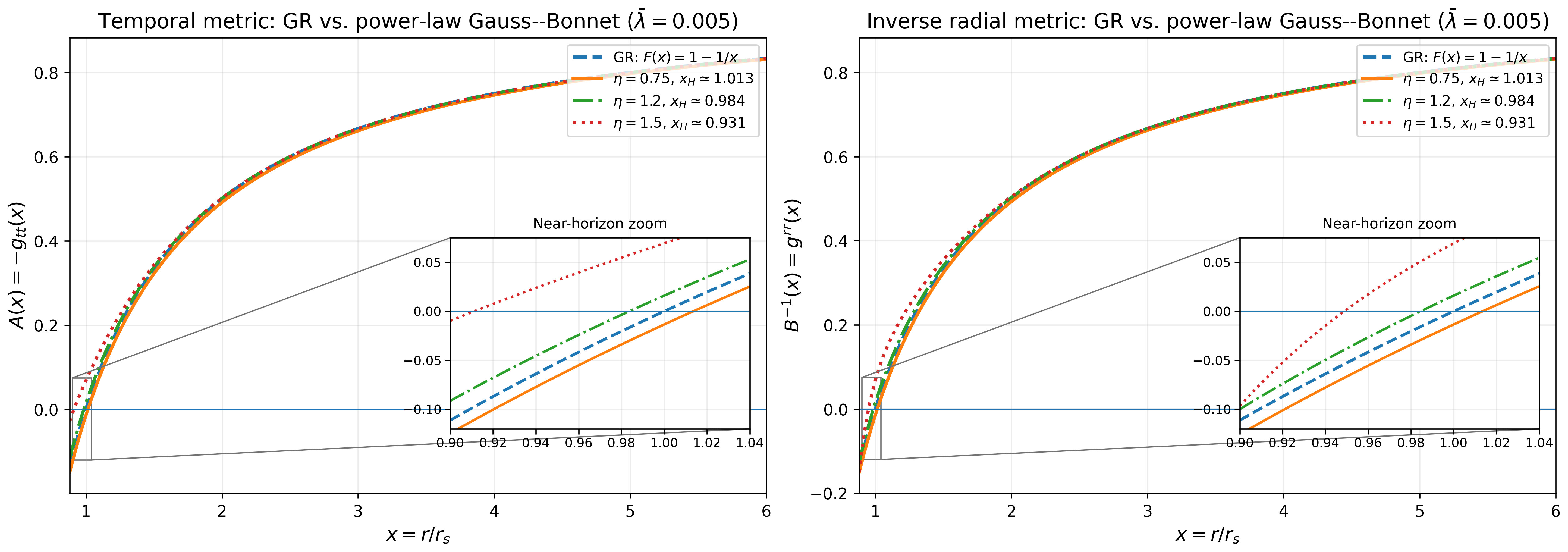}
    \caption{Temporal metric function
    \(\mathcal A=-g_{tt}\) and inverse radial function
    \(\mathcal B^{-1}=g^{rr}\) for
    \(\bar\lambda=5\times10^{-3}\). The insets magnify the common
    displacement of their near-horizon zeros. All displayed values,
    \(\eta=0.75,1.2,1.5\), belong to the fixed-ADM domain
    \(\eta>2/3\), and the local correction vanishes at \(\eta=1\).}
    \label{fig:metric-functions-power-gb}
\end{figure}

Figure~\ref{fig:metric-functions-power-gb} compares the two horizon-defining
metric functions and displays their common displaced zero.

Since \(r\) is the areal radius,
\begin{equation}
    A_H
    =
    4\pi r_H^2
    =
    4\pi r_s^2
    \left[
        1-2\bar\lambda\mathcal C_H(\eta)
    \right]
    +
    \mathcal O(\bar\lambda^2).
    \label{eq:horizon-area-power-gb}
\end{equation}
This area correction will enter the Wald entropy below.

\subsection{ISCO shift}
\label{sec:isco-power-law-gb}

The ISCO considered here is the marginally stable timelike circular orbit
continuously connected to the Schwarzschild value
\(r_{\rm ISCO}^{(0)}=3r_s\). We assume that no disconnected branch of
stable circular orbits enters the perturbative neighborhood under
consideration.

In areal-radius gauge, the general first-order relation is
\begin{align}
    r_{\rm ISCO}
    = {}&3r_s+3\lambda r_s
    \left[18r_s^2h_{tt}''(3r_s)+
        6r_sh_{tt}'(3r_s)+
        h_{tt}(3r_s)\right]+
    \mathcal O(\lambda^2).
\end{align}
Substituting Eq.~\eqref{eq:htt-power-gb} yields
\begin{equation}
    \frac{r_{\rm ISCO}}{r_s}
    =
    3
    -
    \bar\lambda\mathcal C_{\rm ISCO}(\eta)
    +
    \mathcal O(\bar\lambda^2),
    \label{eq:isco-dimensionless-power-gb}
\end{equation}
with
\begin{equation}
        \mathcal C_{\rm ISCO}(\eta)
    =\frac{4^\eta3^{3-5\eta}(\eta-1)
    }{2\eta-1}\left(432\eta^3-       456\eta^2 +118\eta+ 3
    \right).
    \label{eq:isco-coefficient-power-gb}
\end{equation}
Therefore,
\begin{equation}
    \frac{\delta r_{\rm ISCO}}{r_s}
    =
    -\bar\lambda
    \mathcal C_{\rm ISCO}(\eta)
    +
    \mathcal O(\bar\lambda^2).
\end{equation}

\begin{figure}[h!]
    \centering
    \includegraphics[width=0.7\textwidth]
    {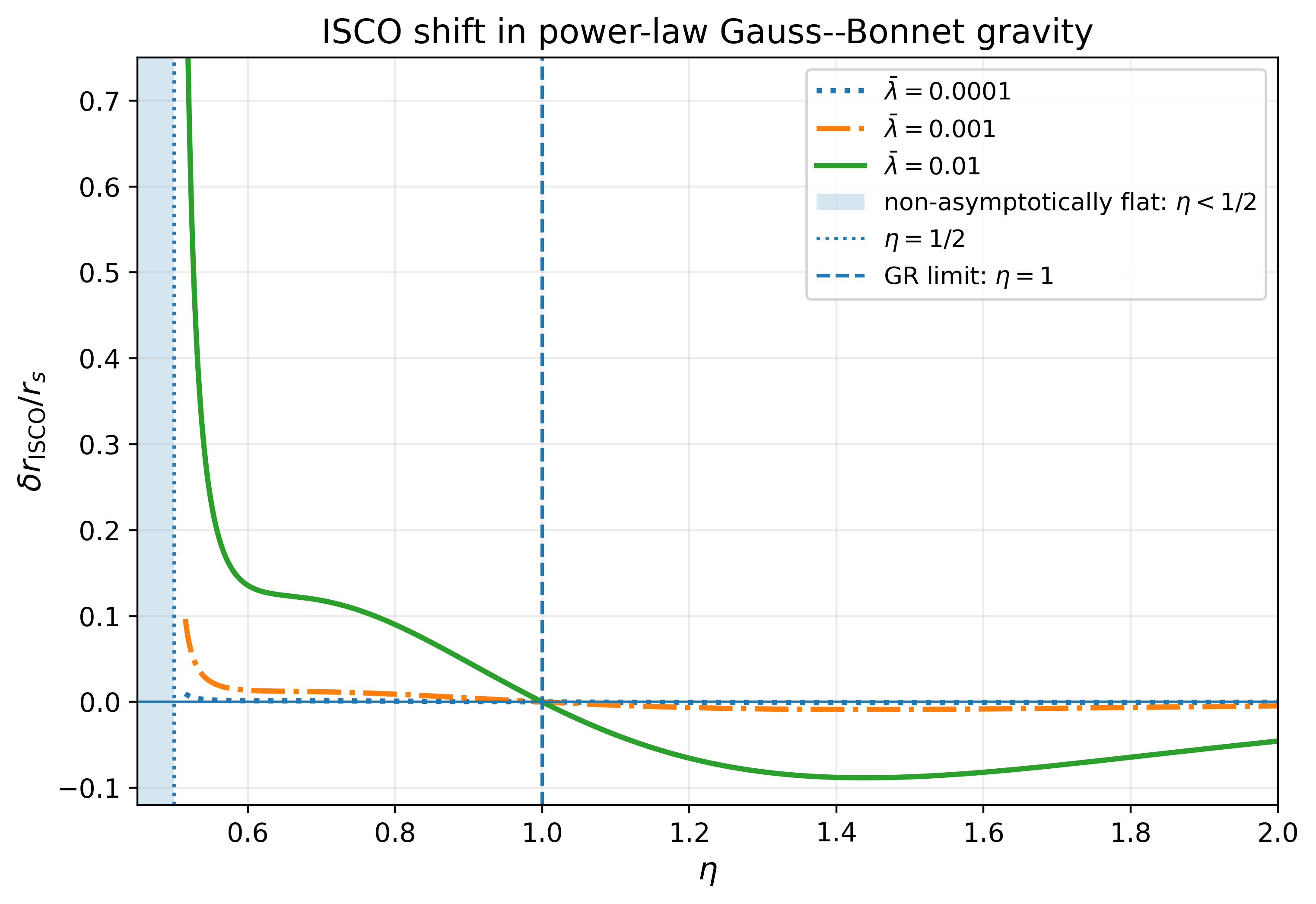}
    \caption{First-order ISCO displacement for representative positive
    values of \(\bar\lambda\). The ISCO moves outward for
    \(1/2<\eta<1\), is unchanged at \(\eta=1\), and moves inward for
    \(\eta>1\). Curves near \(\eta=1/2\), as well as the formal sector
    \(1/2<\eta\leq2/3\), are displayed only to show the behavior of the
    analytic solution and are not used as fixed-ADM predictions.}
    \label{fig:isco-power-gb}
\end{figure}

The cubic polynomial in
Eq.~\eqref{eq:isco-coefficient-power-gb} is positive for
\(\eta>1/2\). The sign of the displacement is therefore controlled by
\(\bar\lambda(\eta-1)\). For positive coupling, the ISCO moves outward
when \(1/2<\eta<1\) and inward when \(\eta>1\), while at \(\eta=1\) the
Schwarzschild value \(r_{\rm ISCO}=3r_s\) is recovered exactly. This
behavior is illustrated in Fig.~\ref{fig:isco-power-gb}.

The ISCO calculation is controlled only when
\begin{equation}
    \frac{
        \left|
            \delta r_{\rm ISCO}
        \right|
    }{
        3r_s
    }
    =
    \frac{
        \left|
            \bar\lambda
            \mathcal C_{\rm ISCO}(\eta)
        \right|
    }{3}
    \ll1.
    \label{eq:isco-validity-power-gb}
\end{equation}

\subsection{Radial epicyclic frequency and periapsis precession}
\label{sec:epicyclic-power-law-gb}

We now specialize the general results of
Sec.~\ref{sec:epicyclic-precession} to the power-law Gauss--Bonnet
solution. For convenience, we introduce
\begin{equation}
    x
    \equiv
    \frac{r_0}{r_s},
    \qquad
    \mathcal D_\eta
    \equiv
    \frac{
        3^{\eta-1}4^\eta(\eta-1)
    }{
        2\eta-1
    }.
    \label{eq:epicyclic-variables-power-gb}
\end{equation}

Substituting Eq.~\eqref{eq:htt-power-gb} into
Eq.~\eqref{eq:orbital-angular-velocity-expanded}, the azimuthal angular frequency becomes
\begin{equation}
    r_s^2\Omega_\phi^2
    =
    \frac{1}{2x^3}
    \left[
        1+\bar\lambda\mathcal W_\eta(x)
    \right]
    +
    \mathcal O(\bar\lambda^2),
    \label{eq:orbital-frequency-dimensionless-power-gb}
\end{equation}
where
\begin{equation}
    \mathcal W_\eta(x)
    =
    -2\mathcal D_\eta x^{3-6\eta}
    \left[
        3\eta(2\eta-1)(4x-3)-1
    \right].
    \label{eq:Weta-definition}
\end{equation}
This function is related directly to the temporal metric perturbation by
\begin{equation}
    \bar\lambda\mathcal W_\eta(x)
    =
    -\lambda r_s x^2 h_{tt}'(xr_s),
    \label{eq:Weta-specialization-power-gb}
\end{equation}
which follows from
Eq.~\eqref{eq:orbital-angular-velocity-expanded}. Therefore,
\begin{equation}
    \frac{
        \Omega_\phi^2-\Omega_{\phi,\rm Schw}^2
    }{
        \Omega_{\phi,\rm Schw}^2
    }
    =
    \bar\lambda\mathcal W_\eta(x)
    +
    \mathcal O(\bar\lambda^2).
    \label{eq:orbital-frequency-fractional-power-gb}
\end{equation}
Spherical symmetry implies
\begin{equation}
    \Omega_\theta=\Omega_\phi,
\end{equation}
so the static geometry produces no nodal precession.

The radial epicyclic frequency satisfies
\begin{equation}
    \frac{\Omega_{\rm rad}^2}{\Omega_\phi^2}
    =
    1-\frac{3}{x}
    +
    \bar\lambda\mathcal Z_\eta(x)
    +
    \mathcal O(\bar\lambda^2),
    \label{eq:radial-frequency-ratio-power-gb}
\end{equation}
where
\begin{equation}
    \mathcal Z_\eta(x)
    =
    3\mathcal D_\eta x^{4-6\eta}
    \left[
        a_\eta
        +
        \frac{2b_\eta}{x}
        +
        \frac{c_\eta}{x^2}
    \right],
    \label{eq:Zeta-dimensionless-power-gb}
\end{equation}
with
\begin{align}
    a_\eta
    &=
    96\eta^3-112\eta^2+32\eta
    =
    16\eta(2\eta-1)(3\eta-2),
    \\
    b_\eta
    &=
    -84\eta^3+112\eta^2-37\eta+1,
    \\
    c_\eta
    &=
    72\eta^3-120\eta^2+52\eta-3.
\end{align}
Direct comparison with Eq.~\eqref{eq:zeta-definition} gives
\begin{equation}
    \lambda\zeta(xr_s,r_s)
    =
    \bar\lambda\mathcal Z_\eta(x),
    \label{eq:zeta-specialization-power-gb}
\end{equation}
which makes explicit the correspondence between the general and
model-specific results.

Combining Eqs.~\eqref{eq:orbital-frequency-dimensionless-power-gb}
and \eqref{eq:radial-frequency-ratio-power-gb}, the absolute radial
frequency is
\begin{align}
    r_s^2\Omega_{\rm rad}^2
    =
    \frac{1}{2x^3}
    \Bigg\{
        1-\frac{3}{x}
        +
        \bar\lambda
        \left[
            \left(
                1-\frac{3}{x}
            \right)
            \mathcal W_\eta(x)
            +
            \mathcal Z_\eta(x)
        \right]
    \Bigg\}
    +
    \mathcal O(\bar\lambda^2).
    \label{eq:radial-frequency-absolute-power-gb}
\end{align}

Marginal stability is determined, at the working perturbative order, by
\begin{equation}
    1-\frac{3}{x_{\rm ISCO}}
    +
    \bar\lambda
    \mathcal Z_\eta(x_{\rm ISCO})
    =
    0.
    \label{eq:marginal-stability-power-gb}
\end{equation}

Expanding about the Schwarzschild value \(x_{\rm ISCO}=3\) gives
\begin{equation}
    x_{\rm ISCO}
    =
    3-3\bar\lambda\mathcal Z_\eta(3)
    +
    \mathcal O(\bar\lambda^2).
    \label{eq:isco-from-epicyclic-power-gb}
\end{equation}
The identity
\begin{equation}
    3\mathcal Z_\eta(3)
    =
    \mathcal C_{\rm ISCO}(\eta)
    \label{eq:Zeta-ISCO-consistency}
\end{equation}
reproduces Eq.~\eqref{eq:isco-dimensionless-power-gb} and provides a
nontrivial check of the metric and orbital calculations.

\begin{figure}[h]
    \centering
    \includegraphics[width=\textwidth]
    {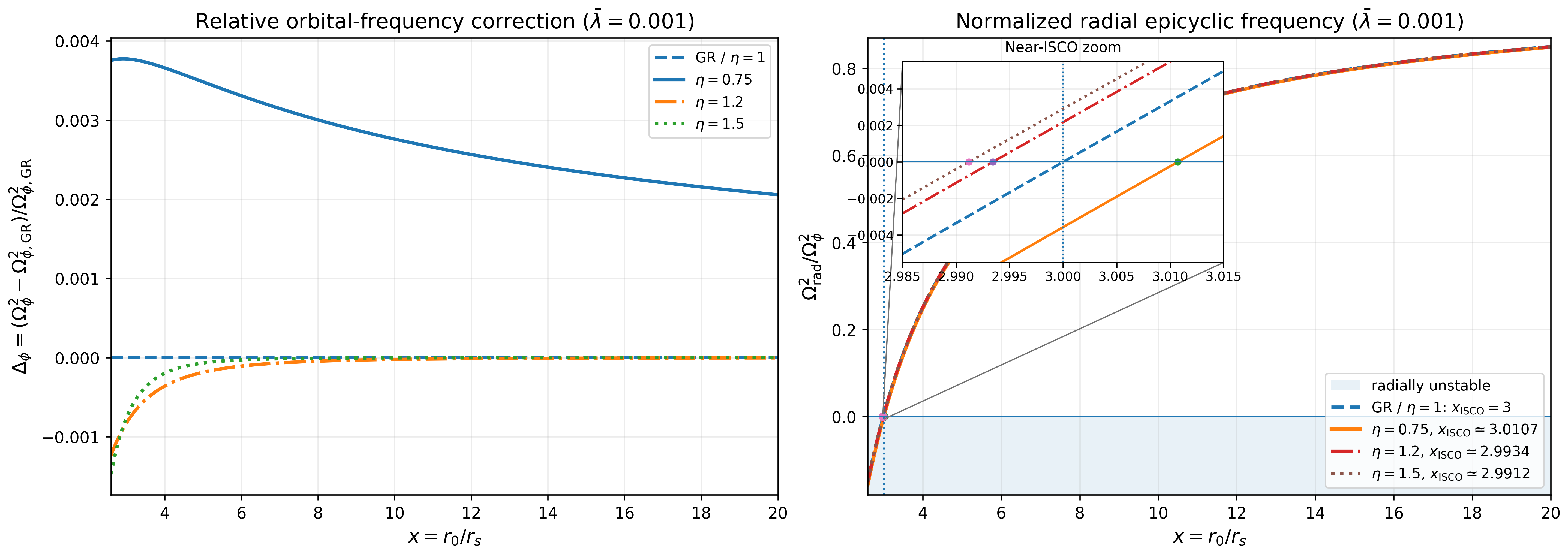}
    \caption{Corrections to the azimuthal and radial epicyclic
    frequencies for \(\bar\lambda=10^{-3}\). The left panel shows the
    fractional correction to \(\Omega_\phi^2\); the right panel shows
    \(\Omega_{\rm rad}^2/\Omega_\phi^2\), with the inset magnifying the
    shifted ISCO. Positive values correspond to radially stable circular
    motion.}
    \label{fig:epicyclic-frequencies-power-gb}
\end{figure}

For the parameter values displayed in
Fig.~\ref{fig:epicyclic-frequencies-power-gb}, with
\(\bar\lambda=10^{-3}\), the first-order ISCO estimates are
\begin{equation}
    x_{\rm ISCO}
    \simeq
    \begin{cases}
        3.0107, & \eta=0.75,\\
        3,      & \eta=1,\\
        2.9935, & \eta=1.2,\\
        2.9913, & \eta=1.5.
    \end{cases}
\end{equation}

For a slightly eccentric orbit about \(r_0\), the periapsis-precession
angle per radial period is
\begin{equation}
    \vartheta
    =
    2\pi
    \left(
        \frac{\Omega_\phi}{\Omega_{\rm rad}}-1
    \right).
    \label{eq:periapsis-definition-power-gb}
\end{equation}
At fixed \(x\), and away from the Schwarzschild ISCO, its first-order
correction is
\begin{equation}
    \Delta\vartheta
    \equiv
    \vartheta-\vartheta_{\rm Schw}
    =
    -\pi\bar\lambda
    \mathcal Z_\eta(x)
    \left(
        1-\frac{3}{x}
    \right)^{-3/2}
    +
    \mathcal O(\bar\lambda^2).
    \label{eq:periapsis-correction-power-gb}
\end{equation}
This expansion requires
\begin{equation}
    \left|
        \bar\lambda\mathcal Z_\eta(x)
    \right|
    \ll
    1-\frac{3}{x}.
    \label{eq:precession-validity-power-gb}
\end{equation}
It necessarily fails arbitrarily close to the Schwarzschild ISCO, where
the unperturbed radial frequency vanishes.

At large \(x\),
\begin{equation}
    \mathcal W_\eta(x)
    \sim
    -8\,\eta\,12^\eta(\eta-1)x^{4-6\eta}.
\end{equation}
The relative orbital-frequency correction therefore decays only for
\(\eta>2/3\). It grows for \(1/2<\eta<2/3\) and approaches a constant
at \(\eta=2/3\). Similarly,
\begin{equation}
    \mathcal Z_\eta(x)
    \sim
    3\mathcal D_\eta a_\eta x^{4-6\eta}.
\end{equation}
Because \(a_\eta\) vanishes at \(\eta=2/3\), the first nonzero radial
correction at that value decays as \(x^{-1}\). Nevertheless, the
\(1/r\) mass mixing described above remains, which is why the fixed-ADM
analysis uses the strict domain \(\eta>2/3\).

\subsection{Photon sphere and critical shadow impact parameter}
\label{sec:photon-shadow-power-law-gb}

We now restrict explicitly to the fixed-ADM domain
\(\eta>2/3\). The photon-sphere radius \(r_{\rm ph}\) is the areal
radius of the outer unstable circular null orbit. By contrast, the shadow
scale is a critical impact parameter in the observer's celestial plane. In
the notation of Sec.~\ref{sec:photon-shadow},
\begin{equation}
    b_{\rm sh}
    \equiv
    b_c,
\end{equation}

The photon-sphere radius is
\begin{equation}
    \frac{r_{\rm ph}}{r_s}
    =
    \frac{3}{2}
    +
    \bar\lambda\mathcal C_{\rm ph}(\eta)
    +
    \mathcal O(\bar\lambda^2),
    \label{eq:photon-sphere-power-gb}
\end{equation}
where
\begin{equation}
    \mathcal C_{\rm ph}(\eta)
    =
    \frac{
        3^{3-5\eta}2^{8\eta-3}
        \eta(1-\eta)(6\eta-1)
    }{
        2\eta-1
    }.
    \label{eq:photon-coefficient-power-gb}
\end{equation}

The critical impact parameter is
\begin{equation}
    \frac{b_{\rm sh}}{r_s}
    =
    \frac{3\sqrt{3}}{2}
    +
    \bar\lambda\mathcal C_{\rm sh}(\eta)
    +
    \mathcal O(\bar\lambda^2),
    \label{eq:shadow-radius-power-gb}
\end{equation}
with
\begin{equation}
    \mathcal C_{\rm sh}(\eta)
    =
    \frac{
        2^{8\eta-4}
        3^{7/2-5\eta}
        (1-\eta)(6\eta+1)
    }{
        2\eta-1
    }.
    \label{eq:shadow-coefficient-power-gb}
\end{equation}

For positive coupling, both \(r_{\rm ph}\) and \(b_{\rm sh}\) increase
when \(2/3<\eta<1\) and decrease when \(\eta>1\). At \(\eta=1\), both
retain their exact Schwarzschild values,
\begin{equation}
    r_{\rm ph}=\frac{3}{2}r_s,
    \qquad
    b_{\rm sh}=\frac{3\sqrt{3}}{2}r_s.
\end{equation}
The corrections vanish exactly because the local geometry is
Schwarzschild for the linear topological interaction.

Although the photon sphere and shadow have the same qualitative sign
change, their coefficients differ:
\begin{equation}
    \frac{
        \mathcal C_{\rm sh}(\eta)
    }{
        \mathcal C_{\rm ph}(\eta)
    }
    =
    \frac{\sqrt{3}}{2}
    \frac{6\eta+1}{\eta(6\eta-1)},
    \qquad
    \eta\neq1,
\end{equation}
with the relation at \(\eta=1\) understood by continuity. This
difference reflects the fact that \(r_{\rm ph}\) is a spacetime areal
radius, whereas \(b_{\rm sh}\) also includes the gravitational redshift
between the photon orbit and the asymptotic observer.

Relative to the Schwarzschild critical impact parameter,
\begin{equation}
    b_{\rm sh}^{\rm Schw}
    =
    \frac{3\sqrt{3}}{2}r_s,
\end{equation}
the fractional correction is
\begin{equation}
    \frac{
        \delta b_{\rm sh}
    }{
        b_{\rm sh}^{\rm Schw}
    }
    =
    \frac{2}{3\sqrt{3}}
    \bar\lambda
    \mathcal C_{\rm sh}(\eta)
    +
    \mathcal O(\bar\lambda^2).
    \label{eq:relative-shadow-shift-power-gb}
\end{equation}

Perturbative control requires
\begin{equation}
    \left|
        \bar\lambda\mathcal C_{\rm ph}
    \right|
    \ll
    \frac{3}{2},
    \qquad
    \left|
        \bar\lambda\mathcal C_{\rm sh}
    \right|
    \ll
    \frac{3\sqrt{3}}{2}.
\end{equation}

\begin{figure}[tbp]
    \centering
    \includegraphics[width=\textwidth]
    {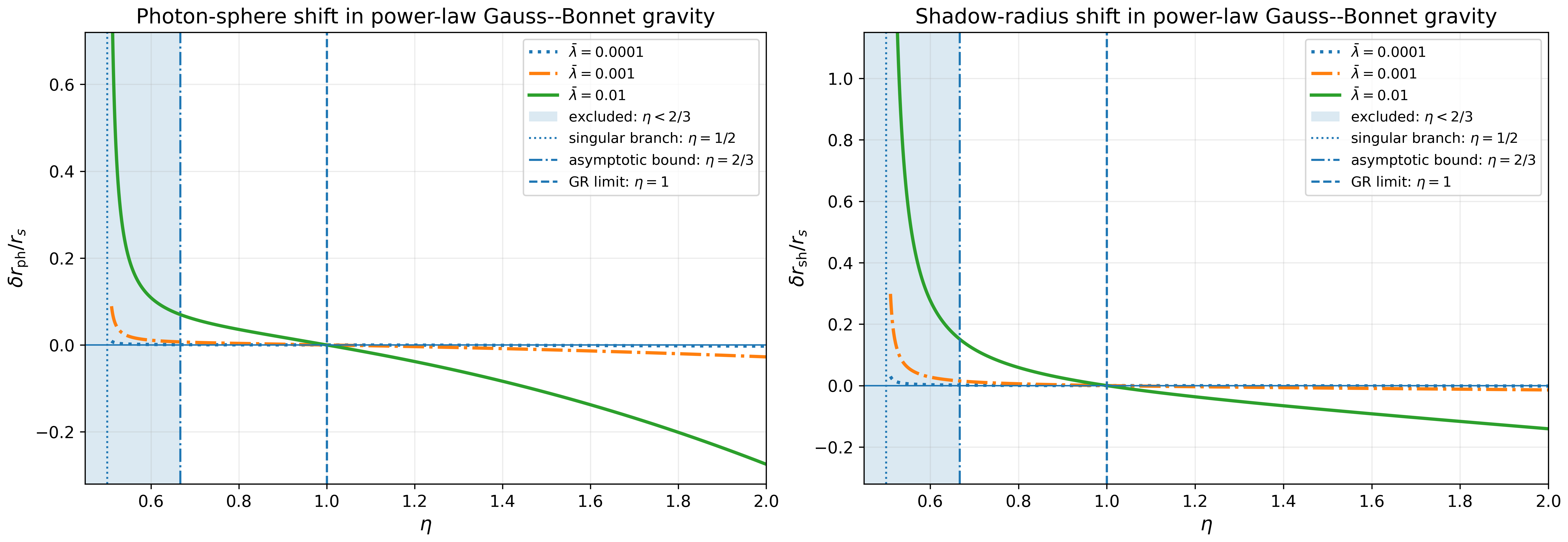}
    \caption{First-order photon-sphere and critical-shadow shifts for
    representative positive values of \(\bar\lambda\). The shaded sector
    \(\eta\leq2/3\) is excluded from the fixed-ADM phenomenological
    analysis. For \(2/3<\eta<1\), both characteristic scales increase;
    for \(\eta>1\), both decrease. The local correction vanishes exactly
    at \(\eta=1\).}
    \label{fig:photon-shadow-power-gb}
\end{figure}

\subsection{EHT-inspired shadow sensitivity and synthetic visualization}
\label{sec:eht-power-law-gb}

The Event Horizon Telescope image of Sgr~A* contains a bright emission
ring with measured angular diameter
\begin{equation}
    d_{\rm ring}
    =
    51.8\pm2.3~\mu{\rm as}
\end{equation}
and is consistent with the expected scale of a Kerr black hole
\cite{EHT2022SgrA,EHT2022SgrAMetric}. The observed emission ring is not
identical to the geometric critical curve, and its position depends on
the plasma, accretion flow, electron thermodynamics, magnetic field, and
radiative transfer.

The present solution is static and spherically symmetric. We therefore
do not perform an EHT likelihood analysis or interpret the result as a
direct observational constraint. Instead, we adopt the conservative
shadow-only consistency criterion
\begin{equation}
    \left|
        \frac{
            \delta b_{\rm sh}
        }{
            b_{\rm sh}^{\rm Schw}
        }
    \right|
    \lesssim
    \epsilon_{\rm EHT},
    \qquad
    \epsilon_{\rm EHT}=0.1.
    \label{eq:eht-shadow-criterion}
\end{equation}
Using Eq.~\eqref{eq:relative-shadow-shift-power-gb}, this gives
\begin{equation}
    |\bar\lambda|_{\rm sh}
    \lesssim
    \frac{
        3\sqrt{3}
    }{
        2
    }
    \frac{
        \epsilon_{\rm EHT}
    }{
        |\mathcal C_{\rm sh}(\eta)|
    }.
    \label{eq:eht-bound-power-gb}
\end{equation}
This curve tests only the assumed ten-percent shadow tolerance. The
region below it is not automatically the complete allowed region of the
theory, nor does it guarantee perturbative control of all other
observables.

A useful simultaneous geometric control criterion is
\begin{equation}
    |\bar\lambda|_{\rm ctrl}
    =
    \epsilon_{\rm pert}
    \min
    \left\{
        \frac{1}{|\mathcal C_H|},
        \frac{3}{|\mathcal C_{\rm ISCO}|},
        \frac{3}{2|\mathcal C_{\rm ph}|},
        \frac{3\sqrt{3}}{2|\mathcal C_{\rm sh}|}
    \right\}.
    \label{eq:simultaneous-geometric-control}
\end{equation}
This expression requires the fractional shifts of the horizon, ISCO,
photon sphere, and shadow to remain below a common tolerance
\(\epsilon_{\rm pert}\). A controlled analysis must additionally satisfy
the local metric-level conditions throughout the radial domain of
interest.

At \(\eta=1\),
\(\mathcal C_{\rm sh}=0\), and the formal shadow-only bound diverges.
The four-dimensional linear Gauss--Bonnet term has no local effect on the
metric or shadow at any order, so the shadow supplies no constraint on its
topological coefficient. Any restriction on that coefficient must come from
considerations other than local metric or geodesic observables. For values close to,
but not exactly equal to, \(\eta=1\), the first-order coefficient becomes
small and higher-order terms may have to be included before drawing a
quantitative conclusion.

\begin{figure}[h]
    \centering
    \includegraphics[width=0.9\textwidth]
    {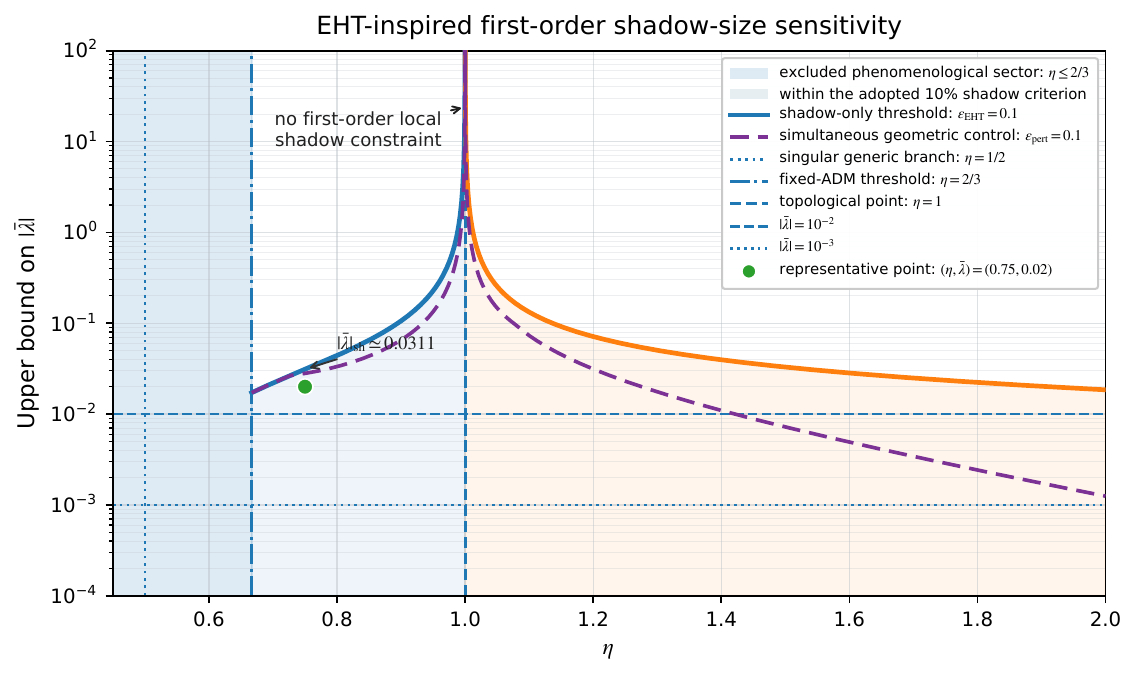}
    \caption{EHT-inspired first-order shadow-size sensitivity for the power-law Gauss--Bonnet model. The solid curve gives the shadow-only ten-percent threshold, with $\epsilon_{\rm EHT}=0.1$; the region below it satisfies this adopted criterion but does not by itself define the complete perturbative or observationally allowed domain. The purple dashed curve shows the simultaneous ten-percent geometric-control threshold. The shaded sector $\eta\leq 2/3$ is excluded from the fixed-ADM phenomenological analysis, while the divergence at $\eta=1$ reflects the exact loss of local metric and shadow sensitivity to the four-dimensional topological Gauss--Bonnet term. The representative point $(\eta,\bar{\lambda})=(0.75,0.02)$ lies below both thresholds; at $\eta=0.75$, $|\bar{\lambda}|_{\rm sh}\simeq0.0311$ and $|\bar{\lambda}|_{\rm ctrl}\simeq0.0280$.}
    \label{fig:eht-bound-power-gb}
\end{figure}

Figure~\ref{fig:eht-bound-power-gb} should consequently be interpreted as
a first-order shadow-sensitivity plot.

At \(\eta=0.75\), the shadow-only ten-percent threshold is
\begin{equation}
    |\bar\lambda|_{\rm sh}
    \simeq
    0.0311.
\end{equation}
The simultaneous ten-percent geometric criterion gives
\begin{equation}
    |\bar\lambda|_{\rm ctrl}
    \simeq
    0.0280.
\end{equation}
We therefore choose
\begin{equation}
    \eta=0.75,
    \qquad
    \bar\lambda=0.02,
    \label{eq:synthetic-shadow-point}
\end{equation}
which remains below both thresholds while producing a visible
displacement.

For this representative point,
\begin{equation}
    \frac{b_{\rm sh}}{r_s}
    \simeq
    2.765,
\end{equation}
whereas
\begin{equation}
    \frac{b_{\rm sh}^{\rm Schw}}{r_s}
    =
    \frac{3\sqrt{3}}{2}
    \simeq
    2.598.
\end{equation}
The relative displacement is
\begin{equation}
    \frac{
        b_{\rm sh}-b_{\rm sh}^{\rm Schw}
    }{
        b_{\rm sh}^{\rm Schw}
    }
    \simeq
    6.43\%.
    \label{eq:synthetic-relative-shadow-shift}
\end{equation}

\begin{figure}[tbp]
    \centering
    \includegraphics[width=\textwidth]
    {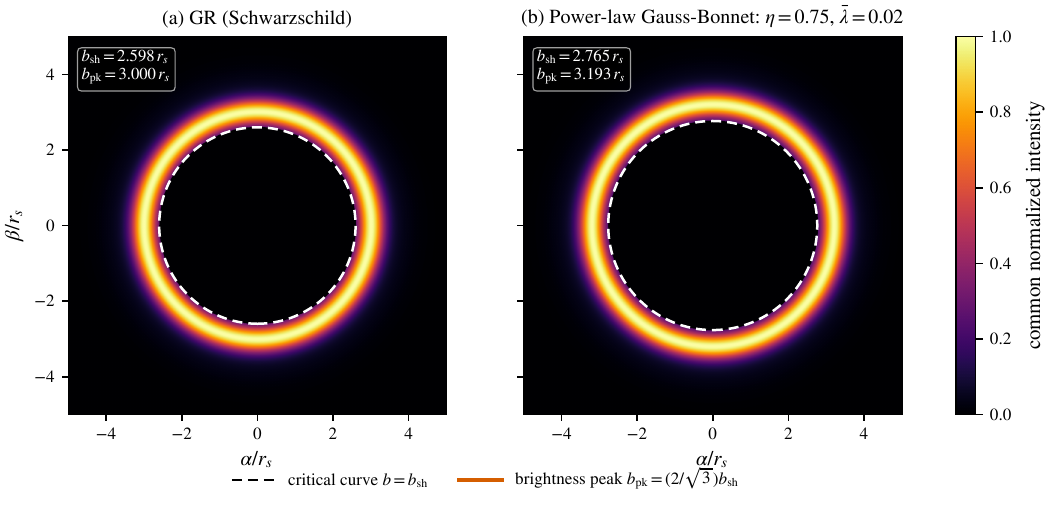}
    \caption{Synthetic image-plane comparison between the Schwarzschild
    geometry and the representative power-law Gauss--Bonnet configuration
    \((\eta,\bar\lambda)=(0.75,0.02)\). The white dashed circumference
    denotes the critical impact parameter \(b=b_{\rm sh}\), with
    \(b_{\rm sh}^{\rm Schw}/r_s=3\sqrt{3}/2\simeq2.598\) and
    \(b_{\rm sh}/r_s\simeq2.765\) in the modified configuration. Both
    panels use the same spatial scale, intensity normalization, and
    schematic radial profile. For visualization only, the brightness
    maximum is prescribed at
    \(b_{\rm pk}=(2/\sqrt{3})b_{\rm sh}\) in both cases; it is not
    identified with the ISCO or with any other emitting orbit. The figure
    illustrates only the first-order displacement of the critical curve
    and should not be interpreted as a ray-tracing or radiative-transfer
    prediction.}
    \label{fig:synthetic-shadow-comparison}
\end{figure}

Figure~\ref{fig:synthetic-shadow-comparison} compares two synthetic
image-plane intensity maps. Both panels use the same spatial scale,
intensity normalization, and radial brightness prescription. The only
geometric circumference shown is the critical curve
\(b=b_{\rm sh}\).

For visualization, the brightness peak is prescribed by
\begin{equation}
    b_{\rm pk}
    =
    \frac{2}{\sqrt{3}}b_{\rm sh}
\end{equation}
in both panels. This common prescription gives
\begin{equation}
    b_{\rm pk}^{\rm Schw}
    =
    3r_s,
    \qquad
    b_{\rm pk}
    \simeq
    3.193r_s.
\end{equation}
The brightness maximum is a schematic image-plane quantity and is not
identified with the ISCO or with any other emitting orbit.

Figures~\ref{fig:eht-bound-power-gb} and
\ref{fig:synthetic-shadow-comparison} therefore serve complementary but
limited purposes. The former displays a shadow-only first-order
sensitivity estimate, whereas the latter visualizes the associated
geometric displacement. A direct comparison with EHT data would require
rotation, a self-consistent accretion model, null-geodesic ray tracing,
and polarized radiative transfer.

\subsection{Wald entropy}
\label{sec:wald-power-law-gb}

We now specialize the general first-order Wald entropy derived in
Sec.~\ref{sec:wald-entropy-general}. The entropy contains two distinct
contributions: the change in the horizon area and the explicit
Noether-charge term generated by the higher-curvature interaction
\cite{Wald1993,IyerWald1994}.

For a static and spherically symmetric perturbation of Schwarzschild,
Eq.~\eqref{eq:wald-first-order-area} gives
\begin{equation}
    S_{\rm W}
    =
    \frac{A_H^{(0)}}{4G}
    +
    \frac{\lambda}{4G}
    \left\{
        A_H^{(1)}
        +
        2A_H^{(0)}
        \left[
            \Psi_R^H
            +
            \frac{4}{r_s^2}\Psi_Y^H
        \right]
    \right\}
    +
    \mathcal O(\lambda^2),
    \label{eq:wald-power-gb-start}
\end{equation}
where
\begin{equation}
    A_H^{(0)}
    =
    4\pi r_s^2,
    \qquad
    R_H^{(0)}=0,
    \qquad
    X_H^{(0)}=0,
    \qquad
    Y_H^{(0)}
    =
    \frac{12}{r_s^4}.
\end{equation}

For the power-law model,
\begin{equation}
    \Psi_{\eta,R}
    =
    2\eta\ell_\lambda^{-2+4\eta}
    R\mathcal G^{\eta-1},
\end{equation}
and
\begin{equation}
    \Psi_{\eta,Y}
    =
    \eta\ell_\lambda^{-2+4\eta}
    \mathcal G^{\eta-1}.
\end{equation}
Thus, on the Schwarzschild horizon,
\begin{equation}
    \Psi_{\eta,R}^H=0,
    \qquad
    \Psi_{\eta,Y}^H
    =
    \eta\ell_\lambda^{-2+4\eta}
    \left(
        \frac{12}{r_s^4}
    \right)^{\eta-1}.
\end{equation}

The area correction obtained from
Eq.~\eqref{eq:horizon-area-power-gb} is
\begin{equation}
    A_H
    =
    A_H^{(0)}
    \left[
        1-2\bar\lambda\mathcal C_H(\eta)
    \right]
    +
    \mathcal O(\bar\lambda^2).
\end{equation}
Its fractional contribution to the entropy is
\begin{equation}
    \frac{
        \delta S_{\rm area}
    }{
        S_{\rm W}^{(0)}
    }
    =
    -2\bar\lambda\mathcal C_H(\eta),
    \qquad
    S_{\rm W}^{(0)}
    =
    \frac{\pi r_s^2}{G}.
\end{equation}

The explicit higher-curvature contribution is
\begin{equation}
    \frac{
        \delta S_{\rm explicit}
    }{
        S_{\rm W}^{(0)}
    }
    =
    8\eta\,12^{\eta-1}\bar\lambda.
\end{equation}
Combining the two terms gives
\begin{align}
    \frac{
        \delta S_{\rm W}
    }{
        S_{\rm W}^{(0)}
    }
    &=
    \bar\lambda
    \left[
        -2\mathcal C_H(\eta)
        +
        8\eta\,12^{\eta-1}
    \right]
    +
    \mathcal O(\bar\lambda^2)
    \nonumber\\
    &=
    \frac{
        8\times12^{\eta-1}
    }{
        2\eta-1
    }
    \bar\lambda
    +
    \mathcal O(\bar\lambda^2).
    \label{eq:wald-relative-power-gb}
\end{align}
Therefore,
\begin{equation}
    S_{\rm W}
    =
    \frac{\pi r_s^2}{G}
    \left[
        1+
        \frac{
            8\times12^{\eta-1}
        }{
            2\eta-1
        }
        \bar\lambda
    \right]
    +
    \mathcal O(\bar\lambda^2).
    \label{eq:wald-entropy-power-gb}
\end{equation}

Equivalently, before introducing \(\bar\lambda\),
\begin{equation}
    \delta S_{\rm W}
    =
    \lambda
    \frac{
        8\pi\,12^{\eta-1}
    }{
        G(2\eta-1)
    }
    \ell_\lambda^{4\eta-2}
    r_s^{4-4\eta}
    +
    \mathcal O(\lambda^2).
\end{equation}

For \(\bar\lambda>0\) and \(\eta>2/3\), the total Wald correction is
positive. When \(2/3<\eta<1\), both the area and explicit
higher-curvature contributions are positive. For \(\eta>1\), the
horizon area decreases, but the positive Noether-charge contribution is
larger in magnitude, and the total entropy still increases. The entropy
therefore cannot be inferred from the area alone.

Perturbative control additionally requires
\begin{equation}
    \left|
        \frac{
            8\times12^{\eta-1}
        }{
            2\eta-1
        }
        \bar\lambda
    \right|
    \ll1.
    \label{eq:wald-validity-power-gb}
\end{equation}
This condition controls the truncated nonlinear branch. At the topological
point \(\eta=1\), the Schwarzschild geometry is exact and the entropy shift
below is exactly linear in the coupling; local geometric control does not
require imposing Eq.~\eqref{eq:wald-validity-power-gb} there.

The linear Gauss--Bonnet case provides an important consistency check.
At \(\eta=1\), the local geometry and horizon area remain exactly
Schwarzschild, but the entropy receives the constant shift
\begin{equation}
    \frac{
        \delta S_{\rm W}
    }{
        S_{\rm W}^{(0)}
    }
    =
    8\bar\lambda,
\end{equation}
or
\begin{equation}
    \delta S_{\rm W}
    =
    \frac{
        8\pi\lambda\ell_\lambda^2
    }{G}.
    \label{eq:wald-topological-shift}
\end{equation}
This mass-independent term is the expected topological contribution of
the four-dimensional linear Gauss--Bonnet density. It changes the
absolute Noether-charge entropy by a constant but does not modify local
field equations, geodesic observables, or differential first-law
relations at fixed horizon topology.

\subsection{Hawking temperature and first-law consistency}
\label{sec:temperature-first-law-power-gb}

For the nonextremal static horizon branch considered above, we assume that
\(h_{tt}\) and \(F_1\) admit finite first derivatives at \(r=r_s\). The
surface gravity associated with the Killing vector
\(\xi^\mu=(\partial/\partial t)^\mu\), normalized at infinity, is
\begin{equation}
    \kappa_H
    =
    \frac{1}{2}
    \left[
        -H'(r_H)
        \left(g^{rr}\right)'(r_H)
    \right]^{1/2}.
    \label{eq:surface-gravity-general}
\end{equation}
Using
\begin{equation}
    H(r)
    =
    -F(r)
    +
    \lambda h_{tt}(r)
    +
    \mathcal O(\lambda^2),
    \qquad
    g^{rr}(r)
    =
    F(r)
    +
    \lambda F_1(r)
    +
    \mathcal O(\lambda^2),
\end{equation}
together with the horizon displacement
\begin{equation}
    r_H
    =
    r_s
    +
    \lambda r_s h_{tt}(r_s)
    +
    \mathcal O(\lambda^2),
\end{equation}
we obtain
\begin{equation}
    \kappa_H
    =
    \frac{1}{2r_s}
    \left\{
        1
        +
        \lambda
        \left[
            -2h_{tt}(r_s)
            -
            \frac{r_s}{2}h_{tt}'(r_s)
            +
            \frac{r_s}{2}F_1'(r_s)
        \right]
        +
        \mathcal O(\lambda^2)
    \right\}.
    \label{eq:surface-gravity-first-order}
\end{equation}

For the power-law Gauss--Bonnet solution in
Eqs.~\eqref{eq:htt-power-gb} and \eqref{eq:F1-power-gb}, the surface
gravity becomes
\begin{equation}
    \kappa_H
    =
    \frac{1}{2r_s}
    \left[
        1
        +
        \bar\lambda\mathcal C_T(\eta)
        +
        \mathcal O(\bar\lambda^2)
    \right],
    \label{eq:surface-gravity-power-gb}
\end{equation}
where
\begin{equation}
    \mathcal C_T(\eta)
    =
    \frac{
        16\times12^{\eta-1}(\eta-1)
    }{
        2\eta-1
    }.
    \label{eq:temperature-coefficient-power-gb}
\end{equation}
Therefore, in units \(\hbar=k_B=1\), the corresponding Hawking
temperature is \cite{Wald:1999vt}
\begin{equation}
T_H=\frac{1}{4\pi r_s}\big[1+
        \bar\lambda\mathcal C_T(\eta)
        +\mathcal O(\bar\lambda^2)\big].
    \label{eq:hawking-temperature-power-gb}
\end{equation}

For positive \(\bar\lambda\), the temperature decreases when
\(2/3<\eta<1\) and increases when \(\eta>1\), consistently with the
corresponding outward and inward horizon displacements, respectively.

The Hawking temperature and Wald entropy provide an independent
first-law consistency check. For this purpose, we restrict the analysis
to the fixed-ADM branch $\eta>2/3$,
for which the homogeneous mass-shift mode has been set to zero and $M=\tfrac{r_s}{2G}$.

We vary the one-parameter family of black-hole solutions at fixed
\(\lambda\), \(\ell_\lambda\), \(\eta\), and horizon topology. Since
\(\bar\lambda\) depends explicitly on \(r_s\), it must not be held fixed
under this variation. From
Eq.~\eqref{eq:lambdabar-definition-power-gb}, one finds
\begin{equation}
    \frac{d\bar\lambda}{dr_s}
    =
    -\frac{4\eta-2}{r_s}\bar\lambda.
    \label{eq:lambdabar-mass-variation}
\end{equation}

Using Eq.~\eqref{eq:wald-entropy-power-gb}, the derivative of the Wald
entropy is
\begin{equation}
    \frac{dS_{\rm W}}{dr_s}
    =
    \frac{2\pi r_s}{G}
    \left[
        1
        -
        \bar\lambda\mathcal C_T(\eta)
        +
        \mathcal O(\bar\lambda^2)
    \right].
    \label{eq:wald-entropy-mass-derivative}
\end{equation}
Combining Eqs.~\eqref{eq:hawking-temperature-power-gb} and
\eqref{eq:wald-entropy-mass-derivative}, we obtain
\begin{equation}
    T_H\frac{dS_{\rm W}}{dr_s}
    =
    \frac{1}{2G}
    \left[
        1+\mathcal O(\bar\lambda^2)
    \right]
    =
    \frac{dM}{dr_s}
    \left[
        1+\mathcal O(\bar\lambda^2)
    \right].
    \label{eq:first-law-power-gb}
\end{equation}
Thus, at fixed couplings and fixed horizon topology, the ordinary first
law \(dM=T_H\,dS_{\rm W}\) is satisfied through first order.

At the topological point \(\eta=1\),
\(\mathcal C_T(1)=0\), and the local geometry is exactly Schwarzschild.
Consequently, the Hawking temperature remains exactly equal to its
Schwarzschild value. The corresponding Gauss--Bonnet entropy shift is
independent of \(r_s\) and therefore drops out of the entropy variation,
consistently leaving the differential first law unchanged.

\section{Conclusions}
\label{sec:conclusions}

We have developed a first-order post-Einsteinian framework for comparing
static and spherically symmetric vacuum black holes in higher-curvature
gravity with a background solution of general relativity. Our main
structural result is that the first-order deformation on the
coupling-analytic branch is controlled by the higher-curvature field
equations evaluated on the GR background. In particular, interactions
constructed solely from \(R\) and \(R_{\mu\nu}\), analytic near the
Ricci-flat point and satisfying \(\Psi(0,0)=0\), do not source vacuum
deformations under the assumptions adopted here. By contrast, suitable
nonlinear Riemann-dependent terms can generate nonvanishing sources
because invariants such as the Kretschmann scalar remain nonzero on
Schwarzschild. This extends the perturbative selection criterion
previously established for Ricci-based theories and identifies nonlinear
Riemann-dependent contributions as a sector capable of producing
coupling-analytic deformations of Schwarzschild geometry. A term linear
in \(Y=R_{\mu\nu\rho\sigma}R^{\mu\nu\rho\sigma}\) is an important
exception: in four dimensions it does not produce a local deformation
because it differs from Ricci-based quadratic terms by the topological
Gauss--Bonnet density.

For a general theory depending on \(R\),
\(X=R_{\mu\nu}R^{\mu\nu}\), and
\(Y=R_{\mu\nu\rho\sigma}R^{\mu\nu\rho\sigma}\), we derived the
first-order field equations in areal-radius gauge and expressed the
horizon, orbital, and optical observables directly in terms of the metric
perturbations. These include the horizon displacement, ISCO,
azimuthal orbital frequency, radial epicyclic frequency, periapsis
advance, photon sphere, and critical shadow impact parameter. We also
derived the first-order Wald entropy, separating the horizon-area
displacement from the explicit Noether-charge contribution. On
Schwarzschild, the \(\Psi_X\) contribution vanishes, whereas the
\(\Psi_R\) and \(\Psi_Y\) terms can remain nonzero, so the Wald entropy
need not be determined by the corrected horizon area alone.

As an explicit application, we considered the power-law Gauss--Bonnet
family \(\Psi_\eta=\ell_\lambda^{-2+4\eta}\mathcal G^\eta\). Since
\(\mathcal G^{(0)}=12r_s^2/r^6\) on Schwarzschild, this interaction
provides a radially varying source for the first-order metric equations.
For noninteger \(\eta\), \(\mathcal G^\eta\) is nonanalytic at
\(\mathcal G=0\), so these models should be interpreted as
phenomenological parametrizations of nonlinear curvature effects. This
nonanalyticity in the curvature argument does not obstruct the
first-order expansion in \(\lambda\) employed here on the exterior
Schwarzschild solution, where \(\mathcal G^{(0)}>0\).

The asymptotic behavior separates the solutions into distinct parameter
regimes. For \(0<\eta<1/2\), the perturbations grow at large radius and
the metric does not approach Schwarzschild. The generic closed-form
solution is not valid at \(\eta=1/2\), which must be treated by solving
the first-order field equations separately. For \(1/2<\eta<2/3\), the
perturbations vanish asymptotically but decay more slowly than the
Schwarzschild \(1/r\) mass term, so the usual conditions required for
the standard ADM interpretation are not satisfied. At \(\eta=2/3\),
the sourced \(1/r\) correction mixes with the homogeneous mass mode. The
fixed-ADM analysis is consequently restricted to \(\eta>2/3\), where
the sourced perturbation decays faster than \(1/r\).

Within the fixed-ADM domain, and wherever the first-order corrections
remain perturbatively controlled, the horizon radius, ISCO radius,
photon-sphere radius, and critical shadow impact parameter display the
same qualitative dependence on the effective coupling. For
\(\bar\lambda>0\), all four quantities increase when
\(2/3<\eta<1\) and decrease when \(\eta>1\); the signs of the shifts
are reversed for \(\bar\lambda<0\). As consistency checks, the ISCO
shift agrees with that obtained independently from the vanishing of the
radial epicyclic frequency, while the perturbed Killing norm and
\(g^{rr}\) vanish at the same corrected horizon radius.

The value \(\eta=1\) provides an exact topological consistency check.
In this case, the interaction reduces to the linear four-dimensional
Gauss--Bonnet density and leaves the local field equations, the
Schwarzschild geometry, and all local geometric and geodesic observables
unchanged. The Wald entropy nevertheless acquires the constant
topological correction
\(\delta S_{\rm W}=8\pi\lambda\ell_\lambda^2/G\). For $\eta\neq1$, the first-order correction is
\(\delta S_{\rm W}/S_{\rm W}^{(0)}
=8\times12^{\eta-1}\bar\lambda/(2\eta-1)+\mathcal O(\bar\lambda^2)\).
Hence, throughout the perturbatively controlled part of the fixed-ADM
domain \(\eta>2/3\), a positive coupling gives a positive first-order
correction to the Wald entropy. For \(\eta>1\), this increase occurs
despite a decrease in the horizon area because the positive
Noether-charge contribution is larger in magnitude than the negative
area correction. We also derived the Hawking temperature from the
surface gravity of the perturbed horizon and verified that the first law
\(dM=T_H\,dS_{\rm W}\) holds through first order on the fixed-ADM branch
under variations at fixed fundamental couplings and fixed horizon
topology.

Finally, we used the correction to the critical shadow impact parameter
to estimate the sensitivity of the geometric shadow scale to
\(\bar\lambda\). This should be interpreted
only as a first-order geometric sensitivity estimate rather than as a
complete observational constraint, since it does not by itself guarantee
perturbative control of the remaining observables or incorporate source
modeling and instrumental uncertainties. At \(\eta=1\), the linear
four-dimensional Gauss--Bonnet term has no local effect on the metric or
shadow, so local geometric observables cannot constrain its topological
coupling.

Several extensions are needed to broaden the scope of these results.
The special case \(\eta=1/2\) should be solved independently, while
\(\eta=2/3\) requires a dedicated asymptotic analysis. A second-order calculation is needed to quantify the
truncation error and determine the domain of validity of the perturbative
expansion, especially near parameter values where first-order
coefficients are enhanced or vanish. Dynamical stability, quasinormal
modes, and gravitational-wave propagation must also be analyzed
independently because the existence of a regular static solution does
not establish its stability in a higher-curvature theory
\cite{DeFeliceSuyamaTanaka2011}. Finally, a direct comparison with
horizon-scale observations will require rotating solutions together with
self-consistent null-geodesic ray tracing, accretion modeling, and
radiative-transfer calculations.

\ack{The authors are grateful to Pedro Bargue{\~n}o, Ángel Rinc{\'o}n, and Ernesto Contreras for stimulating discussions and valuable insights that helped improve this work. J.R. and G.M. acknowledge that this study was financed in part by the Coordenação de Aperfeiçoamento de Pessoal de Nível Superior -- Brasil (CAPES) -- Finance Code 001.}

\roles{All authors contributed to conceptualization, methodology, formal
analysis, validation, and writing (original draft, review, and editing).}

\data{No new data were created or analysed in this study. All results are
based on the analytic calculations presented in the article.}

\bibliographystyle{unsrtnat}
\bibliography{refs}

@article{Wald:1999vt,
  author   = {Wald, Robert M.},
  title    = {{The Thermodynamics of Black Holes}},
  journal  = {Living Reviews in Relativity},
  year     = {2001},
  month    = {Jul},
  volume   = {4},
  number   = {1},
  pages    = {6},
  doi      = {10.12942/lrr-2001-6},
  url      = {https://doi.org/10.12942/lrr-2001-6},
  issn     = {1433-8351}
}

@article{EHT2022SgrAMetric,
  author    = {{Event Horizon Telescope Collaboration} and others},
  title     = {{First Sagittarius A* Event Horizon Telescope Results. VI. Testing the Black Hole Metric}},
  journal   = {The Astrophysical Journal Letters},
  volume    = {930},
  number    = {2},
  pages     = {L17},
  year      = {2022},
  month     = may,
  publisher = {The American Astronomical Society},
  doi       = {10.3847/2041-8213/ac6756},
  url       = {https://doi.org/10.3847/2041-8213/ac6756}
}

@article{NojiriOdintsov2005,
title = {{Modified Gauss–Bonnet theory as gravitational alternative for dark energy}},
journal = {Physics Letters B},
volume = {631},
number = {1},
pages = {1-6},
year = {2005},
issn = {0370-2693},
doi = {https://doi.org/10.1016/j.physletb.2005.10.010},
url = {https://www.sciencedirect.com/science/article/pii/S0370269305014619},
author = {Nojiri, S. and Odintsov, S. D.}
}

@article{DeFeliceTsujikawaCosmology2009,
  author   = {De Felice, Antonio and Tsujikawa, Shinji},
  title    = {{Construction of cosmologically viable $f(G)$ gravity models}},
  journal  = {Physics Letters B},
  year     = {2009},
  volume   = {675},
  number   = {1},
  pages    = {1--8},
  doi      = {10.1016/j.physletb.2009.03.060},
  url      = {https://www.sciencedirect.com/science/article/pii/S0370269309003591},
  issn     = {0370-2693}
}

@book{Poisson2004,
	author    = {Poisson, Eric},
	title     = {{A Relativist's Toolkit: The Mathematics of
	Black-Hole Mechanics}},
	publisher = {Cambridge University Press},
	address   = {Cambridge},
	year      = {2004},
	isbn      = {9780521537803}
}

@book{Chandrasekhar1983,
  author    = {Chandrasekhar, Subrahmanyan},
  title     = {{The Mathematical Theory of Black Holes}},
  publisher = {Oxford University Press},
  address   = {Oxford},
  year      = {1983}
}

@article{BardeenPressTeukolsky1972,
  author  = {Bardeen, James M. and Press, William H. and
             Teukolsky, Saul A.},
  title   = {{Rotating Black Holes: Locally Nonrotating Frames,
             Energy Extraction, and Scalar Synchrotron Radiation}},
  journal = {The Astrophysical Journal},
  volume  = {178},
  pages   = {347--370},
  year    = {1972},
  doi     = {10.1086/151796}
}

@article{Donoghue1994,
  title = {{General relativity as an effective field theory: The leading quantum corrections}},
  author = {Donoghue, John F.},
  journal = {Phys. Rev. D},
  volume = {50},
  issue = {6},
  pages = {3874--3888},
  numpages = {0},
  year = {1994},
  month = {Sep},
  publisher = {American Physical Society},
  doi = {10.1103/PhysRevD.50.3874},
  url = {https://link.aps.org/doi/10.1103/PhysRevD.50.3874}
}

@book{BirrellDavies1982,
  title     = {{Quantum Fields in Curved Space}},
  author    = {Birrell, N. D. and Davies, P. C. W.},
  series    = {Cambridge Monographs on Mathematical Physics},
  publisher = {Cambridge University Press},
  address   = {Cambridge},
  year      = {1982},
  doi       = {10.1017/CBO9780511622632},
  isbn      = {9780521278584}
}

@article{NojiriOdintsov2011,
title = {{Unified cosmic history in modified gravity: From F(R) theory to Lorentz non-invariant models}},
journal = {Physics Reports},
volume = {505},
number = {2},
pages = {59-144},
year = {2011},
issn = {0370-1573},
doi = {https://doi.org/10.1016/j.physrep.2011.04.001},
url = {https://www.sciencedirect.com/science/article/pii/S0370157311001335},
author = {Nojiri, S. and Odintsov, S. D.}
}

@article{Molano2025PRD,
  title = {{Cosmological gauge invariant perturbation theory in $f(R)$ theories of gravity}},
  author = {Molano, Daniel and Villalba, Fabi\'an Dar\'{\i}o and Casta\~neda, Leonardo and Bargue\~no, Pedro},
  journal = {Phys. Rev. D},
  volume = {111},
  issue = {2},
  pages = {024045},
  numpages = {19},
  year = {2025},
  month = {Jan},
  publisher = {American Physical Society},
  doi = {10.1103/PhysRevD.111.024045},
  url = {https://link.aps.org/doi/10.1103/PhysRevD.111.024045}
}

@article{Buonanno:1998gg,
  title = {{Effective one-body approach to general relativistic two-body dynamics}},
  author = {Buonanno, A. and Damour, T.},
  journal = {Phys. Rev. D},
  volume = {59},
  issue = {8},
  pages = {084006},
  numpages = {24},
  year = {1999},
  month = {Mar},
  publisher = {American Physical Society},
  doi = {10.1103/PhysRevD.59.084006},
  url = {https://link.aps.org/doi/10.1103/PhysRevD.59.084006}
}

@article{Ori:2000zn,
  title = {{Transition from inspiral to plunge for a compact body in a circular equatorial orbit around a massive, spinning black hole}},
  author = {Ori, Amos and Thorne, Kip S.},
  journal = {Phys. Rev. D},
  volume = {62},
  issue = {12},
  pages = {124022},
  numpages = {8},
  year = {2000},
  month = {Nov},
  publisher = {American Physical Society},
  doi = {10.1103/PhysRevD.62.124022},
  url = {https://link.aps.org/doi/10.1103/PhysRevD.62.124022}
}

@article{Abramowicz:2011xu,
    author = "Abramowicz, Marek A. and Fragile, P. Chris",
    title = "{Foundations of Black Hole Accretion Disk Theory}",
    eprint = "1104.5499",
    archivePrefix = "arXiv",
    primaryClass = "astro-ph.HE",
    reportNumber = "NSF-KITP-12-055",
    doi = "10.12942/lrr-2013-1",
    journal = "Living Rev. Rel.",
    volume = "16",
    pages = "1",
    year = "2013",
    url      = {https://doi.org/10.12942/lrr-2013-1}
}

@article{Perlick:2021aok,
    author = "Perlick, Volker and Tsupko, Oleg Yu.",
    title = "{Calculating black hole shadows: Review of analytical studies}",
    eprint = "2105.07101",
    archivePrefix = "arXiv",
    primaryClass = "gr-qc",
    doi = "10.1016/j.physrep.2021.10.004",
    journal = "Phys. Rept.",
    volume = "947",
    pages = "1--39",
    year = "2022",
    url = {https://www.sciencedirect.com/science/article/pii/S0370157321003811}
}

@article{Fernandes_2022,
doi = {10.1088/1361-6382/ac500a},
url = {https://doi.org/10.1088/1361-6382/ac500a},
year = {2022},
month = {feb},
publisher = {IOP Publishing},
volume = {39},
number = {6},
pages = {063001},
author = {Fernandes, Pedro G S and Carrilho, Pedro and Clifton, Timothy and Mulryne, David J},
title = {{The 4D Einstein–Gauss–Bonnet theory of gravity: a review}},
journal = {Classical and Quantum Gravity}
}

@article{Lovelock,
    author = {Lovelock, David},
    title = {{The Einstein Tensor and Its Generalizations}},
    journal = {Journal of Mathematical Physics},
    volume = {12},
    number = {3},
    pages = {498--501},
    year = {1971},
    month = {03},
    issn = {0022-2488},
    doi = {10.1063/1.1665613},
    url = {https://doi.org/10.1063/1.1665613}
}

@article{Carroll:2004de,
  title = {{Cosmology of generalized modified gravity models}},
  author = {Carroll, Sean M. and De Felice, Antonio and Duvvuri, Vikram and Easson, Damien A. and Trodden, Mark and Turner, Michael S.},
  journal = {Phys. Rev. D},
  volume = {71},
  issue = {6},
  pages = {063513},
  numpages = {11},
  year = {2005},
  month = {Mar},
  publisher = {American Physical Society},
  doi = {10.1103/PhysRevD.71.063513},
  url = {https://link.aps.org/doi/10.1103/PhysRevD.71.063513}
}

@book{Bambi-book,
    author = "Bambi, Cosimo",
    title = "{Introduction to General Relativity. A Course for Undergraduate Students of Physics}",
    doi = "10.1007/978-981-13-1090-4",
    isbn = "978-981-13-1089-8, 978-981-13-1090-4",
    publisher = "Springer",
    address = "Singapore",
    series = "Undergraduate Lecture Notes in Physics",
    year = "2018"
}

@book{Wald:1984rg,
	author    = {Wald, Robert M.},
	title     = {{General Relativity}},
	publisher = {University of Chicago Press},
	address   = {Chicago},
	year      = {1984},
	isbn      = {9780226870335},
	doi       = {10.7208/chicago/9780226870373.001.0001}
}

@book{Green_Schwarz_Witten_2012,
    author = "Green, Michael B. and Schwarz, John H. and Witten, Edward",
    title = "{Superstring Theory Vol. 1}: {25th Anniversary Edition}",
    doi = "10.1017/CBO9781139248563",
    isbn = "978-1-139-53477-2, 978-1-107-02911-8",
    publisher = "Cambridge University Press",
    series = "Cambridge Monographs on Mathematical Physics",
    month = "Nov",
    year = "2012"
}

@book{Ortin_2004,
	author = "Ortin, Tomas",
	title = "{Gravity and Strings}",
	edition = "2nd ed.",
	doi = "10.1017/CBO9781139019750",
	isbn = "978-0-521-76813-9, 978-0-521-76813-9, 978-1-316-23579-9",
	publisher = "Cambridge University Press",
	series = "Cambridge Monographs on Mathematical Physics",
	month = "Jul",
	year = "2015"
}

@article{CLIFTON20121,
title = {{Modified gravity and cosmology}},
journal = {Physics Reports},
volume = {513},
number = {1},
pages = {1-189},
year = {2012},
issn = {0370-1573},
doi = {https://doi.org/10.1016/j.physrep.2012.01.001},
url = {https://www.sciencedirect.com/science/article/pii/S0370157312000105},
author = {Timothy Clifton and Pedro G. Ferreira and Antonio Padilla and Constantinos Skordis}
}

@article{MolanoBargueno2025,
doi = {10.1088/1361-6382/ad9ce1},
url = {https://doi.org/10.1088/1361-6382/ad9ce1},
year = {2025},
publisher = {IOP Publishing},
volume = {42},
number = {2},
pages = {025010},
author = {Molano, Daniel and Bargue{\~n}o, Pedro},
title = {{Exploring perturbative constraints in higher-order curvature gravity theories}},
journal = {Classical and Quantum Gravity}
}

@article{Penrose1965,
author  = {Penrose, R.},
title   = {{Gravitational Collapse and Space-Time Singularities}},
journal = {Physical Review Letters},
volume  = {14},
pages   = {57--59},
year    = {1965},
doi     = {10.1103/PhysRevLett.14.57},
url     = {https://doi.org/10.1103/PhysRevLett.14.57}
}

@book{HawkingEllis1973,
  author    = {Hawking, S. W. and Ellis, G. F. R.},
  title     = {{The Large Scale Structure of Space-Time}},
  series    = {Cambridge Monographs on Mathematical Physics},
  publisher = {Cambridge University Press},
  address   = {Cambridge},
  year      = {1973},
  doi       = {10.1017/CBO9780511524646}
}

@article{tHooftVeltman1974,
author  = {{'t Hooft}, G. and Veltman, M.},
title   = {{One-loop divergences in the theory of gravitation}},
journal = {Annales de l'Institut Henri Poincar{\'e} A},
volume  = {20},
pages   = {69--94},
year    = {1974}
}

@article{Weinberg1989,
author  = {Weinberg, S.},
title   = {{The Cosmological Constant Problem}},
journal = {Reviews of Modern Physics},
volume  = {61},
pages   = {1--23},
year    = {1989},
doi     = {10.1103/RevModPhys.61.1},
url     = {https://doi.org/10.1103/RevModPhys.61.1}
}

@article{CapozzielloDeLaurentis2011,
author  = {Capozziello, S. and De Laurentis, M.},
title   = {{Extended Theories of Gravity}},
journal = {Physics Reports},
volume  = {509},
pages   = {167--321},
year    = {2011},
doi     = {10.1016/j.physrep.2011.09.003},
url     = {https://doi.org/10.1016/j.physrep.2011.09.003}
}

@article{SotiriouFaraoni2010,
author  = {Sotiriou, T. P. and Faraoni, V.},
title   = {{{$f(R)$} Theories of Gravity}},
journal = {Reviews of Modern Physics},
volume  = {82},
pages   = {451--497},
year    = {2010},
doi     = {10.1103/RevModPhys.82.451},
url     = {https://doi.org/10.1103/RevModPhys.82.451}
}

@article{DeFeliceTsujikawa2010,
author  = {De Felice, A. and Tsujikawa, S.},
title   = {{{$f(R)$} Theories}},
journal = {Living Reviews in Relativity},
volume  = {13},
pages   = {3},
year    = {2010},
doi     = {10.12942/lrr-2010-3},
url     = {https://doi.org/10.12942/lrr-2010-3}
}

@book{Buchbinder1992,
author    = {Buchbinder, I. L. and Odintsov, S. D. and Shapiro, I. L.},
title     = {Effective Action in Quantum Gravity},
publisher = {Institute of Physics Publishing},
address   = {Bristol},
year      = {1992}
}

@article{Stelle1977,
author  = {Stelle, K. S.},
title   = {{Renormalization of Higher-Derivative Quantum Gravity}},
journal = {Physical Review D},
volume  = {16},
pages   = {953--969},
year    = {1977},
doi     = {10.1103/PhysRevD.16.953},
url     = {https://doi.org/10.1103/PhysRevD.16.953}
}

@article{Starobinsky1980,
author  = {Starobinsky, A. A.},
title   = {{A New Type of Isotropic Cosmological Models Without Singularity}},
journal = {Physics Letters B},
volume  = {91},
pages   = {99--102},
year    = {1980},
doi     = {10.1016/0370-2693(80)90670-X},
url     = {https://doi.org/10.1016/0370-2693(80)90670-X}
}

@article{NojiriOdintsovOikonomou2017,
author  = {Nojiri, S. and Odintsov, S. D. and Oikonomou, V. K.},
title   = {Modified Gravity Theories on a Nutshell: Inflation, Bounce and Late-Time Evolution},
journal = {Physics Reports},
volume  = {692},
pages   = {1--104},
year    = {2017},
doi     = {10.1016/j.physrep.2017.06.001},
url     = {https://doi.org/10.1016/j.physrep.2017.06.001}
}

@article{Lanczos1938,
author  = {Lanczos, C.},
title   = {{A Remarkable Property of the Riemann-Christoffel Tensor in Four Dimensions}},
journal = {Annals of Mathematics},
volume  = {39},
pages   = {842--850},
year    = {1938},
doi     = {10.2307/1968467},
url     = {https://doi.org/10.2307/1968467}
}

@article{Zwiebach1985,
author  = {Zwiebach, B.},
title   = {{Curvature Squared Terms and String Theories}},
journal = {Physics Letters B},
volume  = {156},
pages   = {315--317},
year    = {1985},
doi     = {10.1016/0370-2693(85)91616-8},
url     = {https://doi.org/10.1016/0370-2693(85)91616-8}
}

@article{DeFeliceTsujikawaSolarSystem2009,
  title = {{Solar system constraints on $f(\mathcal{G})$ gravity models}},
  author = {De Felice, Antonio and Tsujikawa, Shinji},
  journal = {Phys. Rev. D},
  volume = {80},
  issue = {6},
  pages = {063516},
  numpages = {15},
  year = {2009},
  month = {Sep},
  publisher = {American Physical Society},
  doi = {10.1103/PhysRevD.80.063516},
  url = {https://link.aps.org/doi/10.1103/PhysRevD.80.063516}
}

@article{Bruni1997,
author  = {Bruni, M. and Matarrese, S. and Mollerach, S. and Sonego, S.},
title   = {{Perturbations of Space-Time: Gauge Transformations and Gauge Invariance at Second Order and Beyond}},
journal = {Classical and Quantum Gravity},
volume  = {14},
pages   = {2585--2606},
year    = {1997},
doi     = {10.1088/0264-9381/14/9/014},
url     = {https://doi.org/10.1088/0264-9381/14/9/014}
}

@article{StewartWalker1974,
author  = {Stewart, J. M. and Walker, M.},
title   = {Perturbations of Space-Times in General Relativity},
journal = {Proceedings of the Royal Society of London A},
volume  = {341},
pages   = {49--74},
year    = {1974},
doi     = {10.1098/rspa.1974.0172},
url     = {https://doi.org/10.1098/rspa.1974.0172}
}

@article{Wald1993,
author  = {Wald, R. M.},
title   = {{Black Hole Entropy Is the Noether Charge}},
journal = {Physical Review D},
volume  = {48},
pages   = {R3427--R3431},
year    = {1993},
doi     = {10.1103/PhysRevD.48.R3427},
url     = {https://doi.org/10.1103/PhysRevD.48.R3427}
}

@article{IyerWald1994,
author  = {Iyer, V. and Wald, R. M.},
title   = {{Some Properties of Noether Charge and a Proposal for Dynamical Black Hole Entropy}},
journal = {Physical Review D},
volume  = {50},
pages   = {846--864},
year    = {1994},
doi     = {10.1103/PhysRevD.50.846},
url     = {https://doi.org/10.1103/PhysRevD.50.846}
}

@article{Will2014,
author  = {Will, C. M.},
title   = {{The Confrontation between General Relativity and Experiment}},
journal = {Living Reviews in Relativity},
volume  = {17},
pages   = {4},
year    = {2014},
doi     = {10.12942/lrr-2014-4},
url     = {https://doi.org/10.12942/lrr-2014-4}
}

@article{Blanchet2024,
  author   = {Blanchet, Luc},
  title    = {{Post-Newtonian theory for gravitational waves}},
  journal  = {Living Reviews in Relativity},
  year     = {2024},
  month    = {Jul},
  volume   = {27},
  number   = {1},
  pages    = {4},
  doi      = {10.1007/s41114-024-00050-z},
  url      = {https://doi.org/10.1007/s41114-024-00050-z},
  issn     = {1433-8351}
}

@article{Abbott2016,
  title = {{Observation of Gravitational Waves from a Binary Black Hole Merger}},
  author = {Abbott, B. P. and others},
  collaboration = {LIGO Scientific Collaboration and Virgo Collaboration},
  journal = {Phys. Rev. Lett.},
  volume = {116},
  issue = {6},
  pages = {061102},
  numpages = {16},
  year = {2016},
  month = {Feb},
  publisher = {American Physical Society},
  doi = {10.1103/PhysRevLett.116.061102},
  url = {https://link.aps.org/doi/10.1103/PhysRevLett.116.061102}
}

@article{EHT2019M87,
author  = {{Event Horizon Telescope Collaboration}},
title   = {{First {M87} Event Horizon Telescope Results. {I}. {T}he Shadow of the Supermassive Black Hole}},
journal = {The Astrophysical Journal Letters},
volume  = {875},
pages   = {L1},
year    = {2019},
doi     = {10.3847/2041-8213/ab0ec7},
url     = {https://doi.org/10.3847/2041-8213/ab0ec7}
}

@article{EHT2022SgrA,
  author    = {{Event Horizon Telescope Collaboration} and others},
  title     = {{First Sagittarius A* Event Horizon Telescope Results. I. The Shadow of the Supermassive Black Hole in the Center of the Milky Way}},
  journal   = {The Astrophysical Journal Letters},
  volume    = {930},
  number    = {2},
  pages     = {L12},
  year      = {2022},
  month     = may,
  publisher = {The American Astronomical Society},
  doi       = {10.3847/2041-8213/ac6674},
  url       = {https://doi.org/10.3847/2041-8213/ac6674}
}

@article{MolanoVillalbaCastanedaBargueno2020,
author  = {Molano, D. and Villalba, F. D. and Casta{\~n}eda, L. and Bargue{\~n}o, P.},
title   = {{On Perturbative Constraints for Vacuum {$f(R)$} Gravity}},
journal = {Classical and Quantum Gravity},
volume  = {37},
pages   = {145006},
year    = {2020},
doi     = {10.1088/1361-6382/ab8051},
url     = {https://doi.org/10.1088/1361-6382/ab8051},
eprint  = {2001.10683},
archivePrefix = {arXiv},
primaryClass  = {gr-qc}
}

@article{LuPerkinsPopeStelle2015,
  title = {{Black Holes in Higher Derivative Gravity}},
  author = {L\"u, H. and Perkins, A. and Pope, C. N. and Stelle, K. S.},
  journal = {Phys. Rev. Lett.},
  volume = {114},
  issue = {17},
  pages = {171601},
  numpages = {4},
  year = {2015},
  month = {Apr},
  publisher = {American Physical Society},
  doi = {10.1103/PhysRevLett.114.171601},
  url = {https://link.aps.org/doi/10.1103/PhysRevLett.114.171601}
}

@article{DeFeliceSuyamaTanaka2011,
  title = {{Stability of Schwarzschild-like solutions in $f(R,\mathcal{G})$ gravity models}},
  author = {De Felice, Antonio and Suyama, Teruaki and Tanaka, Takahiro},
  journal = {Phys. Rev. D},
  volume = {83},
  issue = {10},
  pages = {104035},
  numpages = {12},
  year = {2011},
  month = {May},
  publisher = {American Physical Society},
  doi = {10.1103/PhysRevD.83.104035},
  url = {https://link.aps.org/doi/10.1103/PhysRevD.83.104035}
}

@article{MyrzakulovSebastianiZerbini2013,
  author   = {Myrzakulov, R. and Sebastiani, L. and Zerbini, S.},
  title    = {{Topological static spherically symmetric vacuum solutions in gravity $\mathcal{F}(R,G)$}},
  journal  = {General Relativity and Gravitation},
  year     = {2013},
  month    = {Mar},
  volume   = {45},
  number   = {3},
  pages    = {675--690},
  doi      = {10.1007/s10714-012-1493-6},
  url      = {https://doi.org/10.1007/s10714-012-1493-6},
  issn     = {1572-9532}
}

@misc{Nashed2026,
      title={{Photon Sphere and Shadow of a Perturbative Black Hole in $f(R,\mathcal{G})$ Gravity}}, 
      author={G. G. L. Nashed},
      year={2026},
      eprint={2605.10992},
      archivePrefix={arXiv},
      primaryClass={gr-qc},
      url={https://arxiv.org/abs/2605.10992} 
}

@article{BertottiIessTortora2003,
  author   = {Bertotti, B. and Iess, L. and Tortora, P.},
  title    = {{A test of general relativity using radio links with the Cassini spacecraft}},
  journal  = {Nature},
  year     = {2003},
  month    = {Sep},
  volume   = {425},
  number   = {6956},
  pages    = {374--376},
  doi      = {10.1038/nature01997},
  url      = {https://doi.org/10.1038/nature01997},
  issn     = {1476-4687}
}

@article{KramerEtAl2021,
  title = {{Strong-Field Gravity Tests with the Double Pulsar}},
  author = {Kramer, M. and others},
  journal = {Phys. Rev. X},
  volume = {11},
  issue = {4},
  pages = {041050},
  numpages = {53},
  year = {2021},
  month = {Dec},
  publisher = {American Physical Society},
  doi = {10.1103/PhysRevX.11.041050},
  url = {https://link.aps.org/doi/10.1103/PhysRevX.11.041050}
}

@article{Berry2011,
	title = {{Linearized $f(R)$ gravity: Gravitational radiation and Solar System tests}},
	author = {Berry, Christopher P. L. and Gair, Jonathan R.},
	journal = {Phys. Rev. D},
	volume = {83},
	issue = {10},
	pages = {104022},
	numpages = {19},
	year = {2011},
	month = {May},
	publisher = {American Physical Society},
	doi = {10.1103/PhysRevD.83.104022},
	url = {https://link.aps.org/doi/10.1103/PhysRevD.83.104022},
	note = {Erratum: Phys. Rev. D 85, 089906 (2012)}
}

\end{document}